\documentclass[fleqn,10pt]{wlscirep}
\usepackage[utf8]{inputenc}
\usepackage[T1]{fontenc}

\usepackage{algorithm}
\usepackage{algpseudocode}
\usepackage{subfig}

\title{Cyber Physical Games}

\author{Warisa Sritriratanarak, Paulo Garcia}
\affil{International School of Engineering, Chulalongkorn University, Bangkok, Thailand}

\affil{\{warisa.s,paulo.g\}@chula.ac.th}


\begin{abstract}
We describe a formulation of multi-agents operating within a Cyber-Physical System, resulting in collaborative or adversarial games. We show that the non-determinism inherent in the communication medium between agents and the underlying physical environment gives rise to environment evolution that is a probabilistic function of agents' strategies. We name these emergent properties Cyber Physical Games and study its properties. We present an algorithmic model that determines the most likely system evolution, approximating Cyber Physical Games through Probabilistic Finite State Automata, and evaluate it on collaborative and adversarial versions of the Iterated Boolean Game, comparing theoretical results with simulated ones. Results support the validity of the proposed model, and suggest several required research directions to continue evolving our understanding of Cyber Physical System, as well as how to best design agents that must operate within such environments.  
\end{abstract}
\begin{document}

\flushbottom
\maketitle
%
%
\thispagestyle{empty}

\section{Introduction}
Cyber Physical Systems (CPSs) \cite{jazdi2014cyber} are characterized as the integration of concurrent embedded systems (the "cyber" components) , each executing a local control loop \cite{xin2015cyber}, in a shared physical environment. Whilst the Internet of Things (IoT) \cite{rose2015internet} describes distributed systems that are capable of internet communication, exposing data and control access points, CPSs exist at a higher level of abstraction. A system of systems is meaningfully described as a Cyber Physical System if and only if the collective behavior of its constituent systems exhibits properties of interest \cite{singh2021emergence}, and each system can thus influence the others (either directly, or indirectly by affecting relevant properties). A CPS may be designed holistically, in a top-down approach \cite{luckeneder2018systematic}; or, incrementally, in a bottom-up approach \cite{lee2006cyber}, depending on the properties of the physical environment dictated by the domain of application. For example, industrial automation environments \cite{leitao2016industrial} may be designed completely top-down, starting from functional specifications which are iteratively refined into a concurrent system design. Networks of autonomous automobile systems \cite{nanda2019internet}, on the other hand, must necessarily exhibit some degree of bottom-up instantiation, as autonomous vehicles enter and leave the macro system. Cyber-system to cyber-system communication is often prevalent (typically, on top of some IoT architecture), but not strictly necessary.

\par Because the cyber-components of a CPS may have substantial degrees of computational power, significant effect on the underlying shared physical system, and well-specified local goals, it is often desirable to formalize them as \textit{agents} \cite{leitao2016smart}. Thus, a natural next step is to formalize the macro behavior of a CPS as a \textit{game}, either adversarial or collaborative. A significant body or work on distributed games exists: notably, the work of Gutierrez \textit{et al} \cite{GUTIERREZ2023103806, gutierrez2021equilibrium, gutierrez2017reasoning}, which has developed a temporal logic to reason about distributed games strategies.

\par Because CPSs are real instances of distributed systems, and operate on top of communication networks and protocols without strict hard-real time guarantees, the ordering of events obeys Lamport's rules \cite{lamport2019time}: they are not necessarily perceived in the same partial order by any frame of reference in the system (i.e., from the underlying physical system's frame of reference, or by the internal frame of reference of each cyber system). Whilst causality can never be broken, a cyber system may, e.g., effect two actions at distinct times that affect the physical system in the opposite order than expected. Thus, the formalization of CPSs as games must account for relative orderings and possible consensus (or lack thereof) within the system.

\par Thus, this paper explores Cyber Physical Games: scenarios where cyber agents act rationally in pursuit of their own goals, in adversarial and/or collaborative contexts, in the presence (and knowledge of) unpredictable and heterogeneous (across frames of reference) ordering of events. Specifically, this paper offers the following contributions:

\begin{itemize}
    \item We present a formal description of Cyber Physical Games, and how they arise from emergent properties of multi-agent Cyber Physical Systems. We further specify the behavior of collaborative and adversarial Cyber Physical Games using examples from the classical Iterated Boolean Game.
  \item We show that there exist collaborative Cyber Physical Games where collaboration can diminish payoffs for both agents, if the non-deterministic properties of CPSs are not taken into account. Conversely, we show that employing knowledge of such properties can improve payoffs for both agents.
  \item We show that adversarial strategies can (and should) take into account the non-deterministic properties of CPSs, as these greatly influence the evolution of adversarial games.
  \item We show that multiple agents operating within a Cyber Physical System give rise to behavior that can be modeled as probabilistic finite state automata.
  \item We introduce an algorithm for determining the most likely sequence of events (i.e., agents' payoffs) in a game ruled by a probabilistic finite state automata, and show this algorithm can be used to predict system evolution. We further discuss how this model can be used in the future to guide the design of agents operating in CPSs. 
\end{itemize}

This paper is organized as follows: we begin some required background (Section \ref{sec:background}), pointing readers to the relevant literature to understand the remainder of the paper. We then present a description of CPSs (Section \ref{sec:cps}) that underpins our game formalism, which we use to explore the execution of games when agents are in collaborative and adversarial scenarios. We then explore the modeling of game strategies as probabilistic finite state automate (Section \ref{sec:model}), describing an algorithm to predict game evolution.  We evaluate these strategies through simulation techniques in Section \ref{sec:experiments}, and describe related work in the field in Section \ref{sec:related}, with Section \ref{sec:conclusions} offering concluding remarks and highlighting the need for future research directions.

\section{Background}\label{sec:background}

\par An agent is a goal-oriented  system, operating in an \textit{environment}; i.e., a set of physical and/or digital objects with various properties and relationships between objects, which contains the agent itself. An agent is capable of acting upon the environment, modifying some (or all) of its properties. A specific combination of properties' values (including of the agent itself) is defined as a \textit{state} $s$ of the environment.  The set of all possible states is referred to as the environment's \textit{state space} . A rational agent will attempt to meet its goal by performing actions that maximize or minimize its \textit{utility function} \cite{russell2010artificial}.
\par Within the context of this work, the environment is a Cyber Physical System \cite{10.1145/3047402}: a super-system that "\textit{combines cyber systems (computational systems such as microprocessors and digital communication networks) with other physical systems (electromechanical, chemical, structural, and biological systems)} \cite{10.1145/2912149}". Of particular interest to us is the communication between the cyber systems and the environment; particularly, the lack of determinism in such a distributed system whose emergent properties give rise to Cyber Physical Games.
\par As first formally described by Lamport in his seminal 1978 paper \cite{lamport2019time}: "\textit{In a distributed system, it is sometimes impossible to
say that one of two events occurred first. The relation "happened before" is therefore only a partial ordering of the events in the system.}". This property gives rise to interesting phenomena in Cyber Physical Games, that introduces additional complexity compared to traditional turn-based \cite{parsons2012game} or distributed \cite{gutierrez2017reasoning} games.

\section{Cyber Physical Games}\label{sec:cps}

\subsection{Modeling Cyber Physical Systems}

Let a CPS be defined as the tuple $\{E,C,M\}$. $E$ denotes the physical shared environment, which is a set of $k$ properties $p_k$ such that $E = \{p_0, p_1, ... p_{k-2}, p_{k-1}\}$. $C$ denotes the set of $n$ cyber agents operating on the environment, such that $C = \{c_0, c_1, ... c_{n-2}, c_{n-1} \}$. The internal behavior of each cyber agent is not relevant for this work; we point interested readers to the body of work on modeling embedded systems \cite{jantsch2003modeling}. $M$ denotes the \textit{message channels} exchanging between the environment $E$ and the cyber agents $C$, such that $M = \{\{m^i_{c_0},m^o_{c_0}\},\{m^i_{c_1},m^o_{c_1}\}, ... \{m^i_{c_{n-2}},m^o_{c_{n-2}}\}, \{m^i_{c_{n-1}},m^o_{c_{n-1}}\}\}$, where $m^i_{c_{x}}$ denotes an input message channel for cyber agent $x$, i.e., sensing the values of the environment, and $m^o_{c_{x}}$ denotes an output message channel from cyber agent $x$, i.e., acting on the environment.
 Messages conceptually represent all possible interactions between each agent and features of the environment, which we notate using the following notation:  $m^i_{c_{x}} \leftarrow p_y$ denotes sensing by the cyber agent of environment property $p_y$; several properties may be sensed at once; e.g., $m^i_{c_{x}} \leftarrow p_y,p_z$. $p_y \leftarrow m^o_{c_{x}} = z$ denotes updating property $p_y$ by agent $c_x$, with a value of $z$.

 Let $M^o$ denote the set of all output message channels (acting on the environment) and $M^i$ denote the set of all input message channels (sensing the environment), such that $M^o \subset M, M^i \subset M, M^o \cap M^i = \emptyset, M^o \cup M^i = M$. To keep notation simpler, instead of adding additional message channels for inter cyber agent communication, we consider the communication link to be a hidden part of the environment $E$, and we notate direct messages with content $s$ from cyber agent $x$ to $y$ through $m^i_{c_{y}}\leftarrow (s)\leftarrow m^o_{c_{x}}$.

\par Whilst environment evolution obeys continuous laws (typically, differential equations), it is convenient for our purposes to model it discretely. Let $E_{t_E}$ denote the state (i.e., the combination of specific values of all its properties) of the environment at time $t_E$, where the subscript denotes that this is time from the frame of reference of the environment. Similarly, inputs to the environment at time $t_E$ are denoted by $M^o_{t_E}$. Then, the environment is governed by some system function $f: (E, M^o) \rightarrow (E)$, such that $E_{t_E} = f(E_{t_E-1},M^o_{t_E-1})$. Throughout the rest of this paper, unless otherwise stated, time $t$ is assumed to be discrete, regardless of which frame of reference is used.

\par A specific message $m$, either to/from the environment from/to a cyber agent, takes a non-deterministic amount of time to be effected, denoted as transfer latency $l$. For a given sender $c_x$, a message $m$ to the environment $E$ has $l_{c_x}(m) = l_0(m^0_{c_x}) + w, w \sim \mathcal{N}(\mu,\sigma^2)$, where $\mu$ and $\sigma^2$ are the mean and  the variance of the distribution of the random variable $w$, and latency is subject to $l_{c_x}(m) > 0$. A consequence of this, well known in the distributed systems canon \cite{lamport2019part}, is that a sequence of messages $m_1$ emitted at $t_{c_x1}$ and  $m_2$ emitted at $t_{c_x2}$, with $t_{c_x2} > t_{c_x1}$, may be received by the environment as the sequence  $m_2$ received at $t_{E1}$ and  $m_1$ received at $t_{E2}$, with $t_{E2} > t_{E1}$. This holds for an arbitrary number of messages among an arbitrary number of cyber agents such that, even under the assumption that all frames of reference are synchronized, each cyber agent and the environment may perceive the sequence of visible events thus far as any possible permutation of events that does not violate causality. Of course, if our assumption that $w$ is normally distributed holds, some permutations are more likely than others.

\par It it noteworthy that, in purely digital systems such as in the case of distributed computing, consistency, control, synchronization, and consensus algorithms rely on (a) not immediately committing the implications of a message; and/or (b) potentially rolling back the state of the distributed system to a previous one. In a CPS, the function $f: (E, M^o) \rightarrow (E)$ might not allow for such luxuries: i.e., actions upon the environment may be permanent.

\subsection{Types of  Cyber Physical Games}

The goals of agents in CPSs are more likely to be concerned with the \textit{evolution} of the state of the environment, rather than with a specific state. For example, in the case of networks of autonomous automobile systems, the goal is the efficient and safe conclusion of vehicles' routes (potentially, until they leave the environment), rather than any specific configuration of vehicles (albeit these may constitute meaningful sub-goals). Thus, iterated, infinite games such as Iterated Boolean Games \cite{gutierrez2015iterated} are reasonable models to reason about the evolution of CPSs. Unlike traditional turn-based runs, agents in a CPS may issue actions at any time. A game is said to be \textit{collaborative} if agents' goals are not in contradiction (in these cases, cooperation may be beneficial, but this is not a necessary condition). A game is said to be \textit{adversarial} if agents' goals are in contradiction.
\par  A (traditional, non-iterative) Boolean game \cite{bonzon2006boolean} is one where there exists a finite set of propositional variables, typically denoted $V = \{a,b,c,...\}$, with $L_V$ denoting the propositional language that is built from $V$, $\top$ (true), $\bot$ (false), and the typical logical operators. Each agent controls a non-intersecting subset of $V$, and their goal is to satisfy a propositional formula of $L_V$, typically denoted $\phi$ or $\psi$. In Iterated Boolean Games, agents' goals are instead given by formulae of Linear Temporal Logic (LTL) \cite{rozier2011linear}. LTL extends propositional language with the temporal operators $\mathbf{X}$ (\textit{next}), $\mathbf{F}$ (\textit{finally}), $\mathbf{G}$ (\textit{globally}), $\mathbf{U}$ (\textit{until}), $\mathbf{R}$ (\textit{release}), $\mathbf{W}$ (\textit{weak until}), and $\mathbf{M}$ (\textit{strong release}). Furthermore , $\rightarrow$ denotes a sequence from one temporal state to another, sometime in the future, but not necessarily the next one. An agent's \textit{strategy} is its decision function, which performs a given available action, based on current and past input, to attempt to satisfy its goal. Notice that, unlike the scenarios explored in \cite{gutierrez2015iterated}, in Cyber Physical Games an agent's strategy does not induce a unique run, because of the non-deterministic properties of CPSs.
\par Adversarial games arise in CPSs when agents' goals are contradictory. This is the case of, e.g., intelligent adverting systems, modifying what is displayed on screens in a commercial area to direct (indirectly, through psychological influence) consumer traffic towards desired commercial venues. Whilst the morality of such scenarios is dubious at best, their study from a theoretical perspective remains quite interesting. As before, Iterated Boolean Games are a reasonable model; the path from an initial state to a terminal state can represent the evolution of the environment in such systems.
\par Throughout the remainder of this paper, we will implicitly assume that games are limited to two players. Our contributions can scale trivially to an arbitrary number of agents, but two agents provides the simplest case to most clearly describe the behavior of Cyber Physical Games.

\subsubsection{The Collaborative Iterated Boolean Game}\label{sec:evolution}

Consider two cyber agents $c_0$ and $c_1$, and an environment $E$ with properties $\{a,b\} \in V$, where both agents can observe and affect the entire set of properties. However, only one variable update can occur at any point in time (i.e., agents cannot simultaneously update both variables under their control).  This corresponds to environment update function

\begin{equation}
\begin{aligned}
&	E_{t_E} = f(M^o_{t_E-1}) \equiv \\
& \{a,b\}_{t_E} = \{ \begin{cases}x,a\leftarrow m^o_{c_0} = x | m^o_{c_1} = x\\a_{t_E-1},otherwise\end{cases},\begin{cases}x,b\leftarrow m^o_{c_0} = x | m^o_{c_1} = x\\b_{t_E-1},otherwise\end{cases}  \}{t_E-1}
\end{aligned}
\end{equation}

In this case, the environment update function can be obviously derived by the agents' control capabilities; notice this may not be true in the general case (e.g., when the environment has internal state update rules that are not immediately derivable from its inputs).

\par Agent $c_0$'s goal is to satisfy an LTL formula $\phi$, and agent $c_1$'s goal is to satisfy an LTL formula $\psi$. Informally, satisfying $\phi$ and $\psi$ throughout system execution corresponds to positive evolution of the system: e.g., successful movement of autonomous vehicles along the desired paths. System execution traces that do not satisfy either formulae correspond to system execution that does not positively evolve the system. Agents' strategies are to maximize the satisfaction of  $\phi$ and $\psi$ over time.

\par Let the evolution of the environment (a \textit{run}) be denoted using angled brackets: $ \langle E_{tE=0},E_{tE=1},...,E_{tE=k-1},E_{tE=k} \rangle$. Being able to write an evolution that satisfies both $\phi$ and $\psi$ fairly suffices to show the game is collaborative. Two different agent modes affect how the game plays out. An agent is deemed \textit{social} if it is aware of the other agent in the environment and is willing to cooperate with it towards the best mutual payoff, and \textit{unsocial} otherwise. An agent is deemed \textit{optimistic} if it assumes its messages are perceived by the environment in the same order as the agent issues them, and that all frames of reference are synchronized, and \textit{realistic} otherwise. Recall that if agent $c_0$ issues the message sequence $\langle \{a \leftarrow m^o_{c_0} = x_0\}_{tc_0=0}, \{b \leftarrow m^o_{c_0} = x_1\}_{tc_0=1} \rangle$ and agent $c_1$ issues the message sequence $\langle \{a \leftarrow m^o_{c_1} = x_2\}_{tc_1=0}, \{b \leftarrow m^o_{c_1} = x_3\}_{tc_2=1} \rangle$, where $t_{c_0}$ and $t_{c_1}$ are in different frames of reference, the environment may perceive the sequences $\langle \{a \leftarrow m^o_{c_0} = x_0\}_{tE=0}, \{b \leftarrow m^o_{c_0} = x_1\}_{tE=1},\{a \leftarrow m^o_{c_1} = x_2\}_{tE=2}, \{b \leftarrow m^o_{c_1} = x_3\}_{tE=3} \rangle$, $\langle \{\{a \leftarrow m^o_{c_1} = x_2\}_{tE=2}, \{b \leftarrow m^o_{c_1} = x_3\}_{tE=3},a \leftarrow m^o_{c_0} = x_0\}_{tE=0}, \{b \leftarrow m^o_{c_0} = x_1\}_{tE=1} \rangle$, ..., or any of the 24 permutations of the 4 individual messages, assuming all messages arrive sequentially in time. The environment may additionally perceive an arbitrarily large number of sequences, corresponding to messages arriving non-sequentially, with arbitrary delays between messages, as given by $l_{c_x}(m) = l_0(m^0_{c_x}) + w, w \sim \mathcal{N}(\mu,\sigma^2)$. Throughout the rest of this paper, we assume all messages arrive sequentially without any delays: i.e., assuming a model where $l_{c_x}(m) \leq 1, \forall m \in M$, and messages are queued. Exploring the impact of message delays $>1$ is left for future work. For a specific set of $n$ messages $\mathbf{m}_n = \{m_0,m_1, ..., m_{n-1}\}$,  we denote its possible no-delay permutations as $P(\mathbf{m}_n)$, where $\|P(\mathbf{m}_n)\| = n!$. We assume social agents have negotiated the common strategy at the beginning of the game, through inter-agent message passing.

\par \textbf{Unsocial, optimistic agents} care not for collaboration, and assume their actions are immediately, and in the same order, effected on the environment. Thus, agents $c_0$ and $c_1$ issue only the messages that would evolve the environment according to $\phi$ and $\psi$, respectively. \textbf{Unsocial but realistic agents} can observe the state of the environment to ensure the outcome of their actions successfully occurred. In this case, they can ensure that their first state transition successfully occurred, before issuing the second message. \textbf{Social but optimistic agents} collaborate to satisfy the compound state transition sequences that satisfies both $\phi, \psi$ fairly, distributing the work among themselves. \textbf{Social and realistic agents} collaborate, timing their actions to ensure mutual benefit, with agents monitoring the environment to preemptively abort failed strategies, if any.

\subsubsection{The Adversarial Iterated Boolean Game}\label{sec:strategies}

Consider two cyber agents $c_0$ and $c_1$, and an environment $E$ with properties $\{a,b,c\} \in V$, where both agents can observe all variables, but affect non-intersecting subsets of properties. Agent $c_0$ has control over variables $a$ and $c$, and agent $c_1$ has control over variable $b$. As in the collaborative case, only one variable update can occur at any point in time (i.e., agents cannot simultaneously update both variables under their control).  This corresponds to environment update function

\begin{equation}
\begin{aligned}
& 	E_{t_E} = f(M^o_{t_E-1}) \equiv \\
& \{a,b,c\}_{t_E} = \{ \begin{cases}x,a\leftarrow m^o_{c_0} = x \\a_{t_E-1},otherwise\end{cases},\begin{cases}x,b\leftarrow m^o_{c_1} = x\\b_{t_E-1},otherwise\end{cases}, \begin{cases}x,c\leftarrow m^o_{c_0} = x \\c_{t_E-1},otherwise\end{cases}  \}{t_E-1}
\end{aligned}
\end{equation}

\par As before, agent $c_0$'s goal is to satisfy an LTL formula $\phi$, and agent $c_1$'s goal is to satisfy an LTL formula $\psi$. As in the collaborative case, an agent is deemed \textit{social} if it's aware of the other agent operating in the environment and, in the adversarial case, actively tries to minimize the other agent's payoff, or \textit{unsocial} otherwise. An agent is deemed \textit{optimistic} if it assumes its messages are perceived by the environment in the same order as the agent issues them, and that all frames of reference are synchronized, and \textit{realistic} otherwise. We assume each agent is aware of the other agent's LTL formula.

\par In the collaborative case, we assumed a a sequence of agents' messages could be partitioned uniformly, regardless of total sequence size. That assumption does not result in an unmanageable number of scenarios, since both agents were aligned in wanting to return to the same initial state after either a successful or unsuccessful run. In the depicted adversarial case, where agents have (potentially) several options to satisfy their formulae, that permutation assumption explodes the state space of the problem. Thus, let us begin the analysis with a simplification of the partitioning assumption.
\par Assuming both agents issue messages at all points in time within their frames of reference and that their frames of reference are synchronized with each other, the environment processes messages issued by both agents in the same order as issued, but with random ordering within message pairs. I.e., give two messages $m_0$ and $m_1$ emitted by agents $c_0$ and $c_1$ synchronously, the environment may process first $m_0$, first $m_1$, or both at the same time (the latter is equivalent to a traditional turn-based game, where both players must commit to an action at any turn, and both actions are executed concurrently). Because the game is infinite, we construct not a game tree from the current to terminal states, but rather a closed game graph, where payoffs are associated with specific walks within the graph.

\par \textbf{Unsocial, optimistic agents} believe they are alone in operating in the environment, and assume their actions are immediate and permanent; i.e., they do not conceive of environment state change without performing an action (leading to potential mismatch between actual state and their belief of state). An agent  attempts to satisfy its formula, prioritizing the graph sub-run that allows it to do so most quickly. \textbf{Unsocial but realistic agents}, are aware the environment may not respond to their actions responsively; thus, they probe for action effect prior to issuing the next action in a sequence. \textbf{Social, optimistic agents} are aware of the competition. Thus, they attempt to chart the path that maximizes the satisfaction of their formula, whilst minimizing the satisfaction of the adversary's formula. Because they conceive of a responsive environment, they assume an action issued at a given moment in time will have effect immediately, synchronously with the adversary's action, such that the total effect on the environment is the sum of both actions (i.e., as in a classical concurrent game). Traditional \textit{minimaxing} strategies cannot be applied in the absence of terminal game states \cite{wen1994folk}; interested readers may consult Devroye et al \cite{devroye1996random} for meaningful strategies in these scenarios. Since the study of strategy optimality is beyond the scope of this paper, we employ a simple strategy that, whilst not necessarily optimal, suffices to illustrate the main points: each agent assumes its adversary will attempt to either carry out a formula satisfaction sequence, if it's within a formula path state, or reach the nearest root state for a formula satisfaction sequence otherwise. Under that assumption, each agent attempt to steer the state using the same reasoning for its own formula satisfaction (both agents look only one state in the future). \textbf{Social but realistic agents}, employing the same strategic principles as in the previous case, can now take advantage of the non-deterministic behavior of CPSs to probabilistically attempt to satisfy their formula in situations that would be impossible in a deterministic concurrent game. With knowledge that it is possible their actions are effected before the adversary's action (regardless of issue order), or vice-versa, agents can now employ a "best case" strategy that will satisfy their formula if the right action is executed first. Again, we are assuming agents plan one state into the future, and assume that the order of actions issued by both agents at the same time step is random, but the ordering of message sequences per agent is preserved.

\section{Stochastic Strategies for Cyber Physical Games' Payoff }\label{sec:model}

Analyzing payoffs, for a system with a small number of states, can be done by listing all run permutations and determine average payoff for each agent. This, of course, does not scale; the general case is more challenging, and requires realizing that the problem must be formulated as determining the probabilities that certain sequences of state transition occur. Notice that, e.g., the transition from state $\{\overline{a}\overline{b}\overline{c}\}$ to state $\{\overline{a}\overline{b}c\}$ might happen because:

\begin{itemize}
	\item At state $\{\overline{a}\overline{b}\overline{c}\}$, $c_0$ issues action $c \leftarrow m^o_{c_0} = \top$ and $c_1$ issues action $b \leftarrow m^o_{c_1} = \bot$, both interpreted by the environment simultaneously.
	\item At state $\{\overline{a}\overline{b}\overline{c}\}$, $c_0$ issues action $c \leftarrow m^o_{c_0} = \top$ and $c_1$ issues action $b \leftarrow m^o_{c_1} = \top$, but the environment interprets $c_0$'s action first.
	\item At state $\{\overline{a}b\overline{a}\}$, $c_0$ issues action $c \leftarrow m^o_{c_0} = \top$ and $c_1$ issues action $b \leftarrow m^o_{c_1} = \bot$, but the environment interprets $c_1$'s action first.
	\item Etc.
\end{itemize}

Thus, rather than look at state transitions, we must look at the probabilities of transition sequences, in function of the environment's interpretation of each message sequence, and of agents' strategies. Notice that, because there is dependence on sequence, rather than just current state, it is not possible to perfectly model this behavior as a Markov Chain \cite{renault2006value}, which assumes unique dependence on just current state; rather, the problem is intractable (specifically, it is NP-hard in the general case \cite{10.1093/logcom/exs049}).

\par Because we are dealing with instances of real systems obeying (partially) known constraints, rather than abstract mathematical models, we may approximate the behavior by taking advantage of our knowledge of CPSs. Recall that a message $m$ to the environment $E$ has $l_{c_x}(m) = l_0(m^0_{c_x}) + w, w \sim \mathcal{N}(\mu,\sigma^2)$: thus, messages are more likely to be interpreted simultaneously rather than sequentially, and each message is equally likely to be interpreted first \cite{kwak2017central}, with possibilities: $c_0$ action effect, followed by $c_1$ action effect; $c_1$ action effect, followed by $c_0$ action effect; or, simultaneous effect of $c_0$'s and $c_1$'s actions. Let us denote these probabilities
$\mathbb{P}(c_o,c_1)$, $\mathbb{P}(c_1,c_0)$, and
$\mathbb{P}(  c_o \| c_1  )$, respectively, where index $k$ represents the environment's frame of reference, where time progresses twice as fast as in the agents' frame of reference.

\par The conditional probability $\mathbb{P}(t| e_k)$ that a state transition $t: e_k \rightarrow e_{k+1}, e \in E$ occurs, given that we are in state $e_k$, can thus be approximated by the sum of possible sequences that include $t$, weighted by the probabilities of simultaneous or sequential interpretation, normalized so the sum of probabilities of all transitions that begin in $e_k$ add up to 1. Let $I_{sim}^{e_k}$ denote the set of all simultaneous interpretations in $e_k$ that result in $t$ (this set has either 0 or 1 element); $I_{seq^1}^{e_k}$ denote the set of all sequential interpretations that contain $e_k$ as the first state (one interpretation) and $I_{seq^2}^{e_k}$ denote the set of all sequential interpretations that contain $e_k$ as the  intermediate state  (second interpretation), including $t$; and $g$ denote the required scaling factor per state to satisfy probability sums; then:

\begin{equation}
\begin{aligned}
&    \mathbb{P}(t | e_k) =  \frac{\mathbb{P}(c_0\|c_1) \|I_{sim}^{e_k}\| + \mathbb{P}(c_0,c_1)\|I_{seq^1}^{e_k}\|+\mathbb{P}(c_1,c_0) \|I_{seq^2}^{e_k}\|}{g}\\
&    \sum\limits_{i = t \in T_{e_k}}^{} \mathbb{P}(i | e_k)= 1\\
&    \mathbb{P}(c_0\|c_1) + \mathbb{P}(c_0,c_1) + \mathbb{P}(c_1,c_0)= 1
\end{aligned}
\label{eq:probabilities}
\end{equation}

\par For each game, we can construct the state sequence matrix $\mathbf{T}$, where row index $i$ denotes the state at time $k$ and column index $j$ represents the next state at time $k+1$, and entry $t_{i,j}$ denotes the probability of transitioning to state $j$ given current state $i$. The value for each entry  $t_{i,j}$ is given by Equation \ref{eq:probabilities}; e.g.:

\begin{equation}
\begin{aligned}
&    \mathbf{T} =
&\begin{bmatrix}
    			0  & \mathbb{P}(0 \rightarrow 1| 0)  & \hdots & \mathbb{P}(0 \rightarrow n| 0)\\
    			\mathbb{P}(1 \rightarrow 0| 1)	 & 0 &\hdots  & \mathbb{P}(1 \rightarrow n| 1)\\
    			\vdots	& \vdots & \ddots &\vdots \\
    			\mathbb{P}(n \rightarrow 0| n)	 & \mathbb{P}(n \rightarrow 1| n) & \hdots & 0\\
                 \end{bmatrix}
\end{aligned}
\end{equation}

\par Thus, we approximate the behavior of a game graph as a \textit{probabilistic finite automaton (PFA)} \cite{vidal2005probabilistic}, where the probabilities of transitioning between states are a function of agents' strategies. These classes of processes are also sometimes called \textit{generative finite automata} \cite{buttelmann1971generalized}: an automaton stochastically generates a path (a substring, in the traditional formulation as a grammar \cite{cognetta2018incremental}).

\begin{algorithm}
\caption{Most likely paths calculation}\label{alg:path}
\begin{algorithmic}
\Require state sequence matrix \textbf{T}
\State Set of most likely paths $\Pi \gets ()$ \Comment{initially empty}
\ForAll{$i$ rows $\in \mathbf{T}$}
  \State Path $\pi \gets i$  \Comment{Adds initial state to ordered path $\pi$}
  \State Current $\gets i$
  \State Next $\gets Max(j) \in$ row $i$ of \textbf{T}

  \While{Next $\notin \pi$}     \Comment{Terminate when we have found a loop}
        \State $Enqueue(Next,\pi)$  \Comment{Adds next state to ordered path $\pi$}
        \State Current $\gets$ Next
        \State Next $\gets Max(j) \in$ row Current of \textbf{T}
    \EndWhile

    \State Current $\gets FirstElement(\pi)$
    \While{Current $\neq$ Next}     \Comment{Prune non-loop states}
        \State $RemoveFirstElement(\pi)$
        \State Current $\gets FirstElement(\pi)$
    \EndWhile

  \If{$\pi \notin \Pi$}
        \State $\Pi \gets \Pi | \pi$ \Comment{Add current path to set of paths}
    \EndIf
\EndFor
\State
\Return  Set of most likely paths $\Pi$
\end{algorithmic}
\end{algorithm}

\par Our goal is to find the likelihood of generating a path satisfying each formula; we point readers to the 2-part review by Vidal et al \cite{vidal2005probabilistic} of (tractable) closed-form solutions for particular cases of such problems in PFAs. In the general case, no closed form solution exists \cite{chatterjee2010probabilistic}; algorithmically, this can be achieved in polynomial time through Monte-Carlo methods \cite{10.1093/logcom/exs049}. Algorithms for finding particular properties, such as finding the most likely generated string, also exist \cite{de2013computing}.

\par We simplify the problem heuristically, and make use of Equation \ref{eq:probabilities} to identify the most likely closed path. Since there are no idle transitions (i.e., transitions where the source and target states are the same), choosing a single transition per source state must necessarily result in a closed path (through the pigeonhole principle). As long as there exists a unique transition, per source state, with highest probability, this path is identifiable and suffices to predict agents' payoffs. Should a unique transition with highest probability not exist in any given state, multiple equally likely closed paths can be analyzed in the same manner. It is also possible that several most likely paths without any states in common exist; these can be identified in the same way. The procedure for determining the most likely path, assuming a single transition with highest probability exists, is described in Algorithm \ref{alg:path}.

\par The set of most likely paths per PFA can be determined, for each matrix, through Algorithm \ref{alg:path}, for specific values or distributions of $\mathbb{P}(c_0,c_1)$, $\mathbb{P}(c_1,c_0)$, and $\mathbb{P}(c_0\|c_1)$ (these can be determined empirically through CPS profiling).

\section{Experiments and Results}\label{sec:experiments}

Our goal here is to analyze whether our models for the evolution of Cyber Physical Games provide an accurate prediction of behavior. Toward this purpose, we develop a computer simulation of  two examples  (corresponding to the collaborative and adversarial Iterated Boolean Game) and explore the predictive power of the models (particularly, the developed heuristics) for different versions of CPS parameters. All code and data for those wanting to reproduce or extend our results is available here\footnote{\url{https://github.com/paulo-chula/Support-code-for-Cyber-Physical-Games.git}}.

\subsection{Evaluating the Collaborative Game}

\par For the collaborative case, we define that agent $c_0$'s goal is to satisfy a formula $\phi$, and agent $c_1$'s goal is to satisfy a formula $\psi$, given by:

\begin{equation}\label{eq:phi_psi}
\begin{aligned}
	\phi : \{\overline{a}\overline{b}\} \rightarrow \mathbf{F} \{a\overline{b}\} \rightarrow \mathbf{X} \{ab\}\\
	\psi : \{\overline{a}\overline{b}\} \rightarrow \mathbf{X} \{\overline{a}b\} \rightarrow \mathbf{F} \{ab\}
\end{aligned}
\end{equation}

For the collaborative case, if agents do not sense the environment after the initial set up, we simply permute the set of agents' actions per iteration, and apply a the generated sequence to the environment. All permutations are equally likely through the Fisher-Yates method \cite{durstenfeld1964algorithm}. For the cases where agents sense the environment (potentially aborting their action issuing in the case of a failed application of their action), we permute in a staggered fashion, interleaving sets bookmarked by agent sensing.
\par It is unnecessary to list all possible runs, but assuming an initial state of $\{\overline{a}\overline{b}\}$, it is trivial to deduce the two optimal runs that satisfy both $\phi$ and $\psi$ fairly:

\begin{equation}\label{eq:mutual}
\begin{aligned}
&	\langle \{\overline{a}\overline{b}\}_{tE=0}, \{\overline{a}b\}_{tE=1}, \{\overline{a}\overline{b}\}_{tE=2}, \{a\overline{b}\}_{tE=3},\{ab\}_{tE=4} \rangle\\
&	\langle \{\overline{a}\overline{b}\}_{tE=0}, \{\overline{a}b\}_{tE=1}, \{ab\}_{tE=2}, \{a\overline{b}\}_{tE=3},\{ab\}_{tE=4} \rangle\\
\end{aligned}
\end{equation}

That these exist suffices to show the game is collaborative. Notice that the former satisfies both formulae at the same time, whilst the latter satisfies $\psi$ before $\phi$. Continuous evolution requires the run to resume to the initial state $\{\overline{a}\overline{b}\}$ (not depicted). Table \ref{tab:boolean_game_colab} depicts agents' strategies in function of their modes:

\begin{table}[h!]
\centering
\begin{tabular}{l | c|  c }
Modes  & $c_0$ & $c_1$ \\
\hline
    Unsocial, Optimistic    & $\langle \{a \leftarrow  \top\}_{0}, \{b \leftarrow  \top\}_{1} \rangle$  & $\langle \{b \leftarrow  \top\}_{0}, \{a \leftarrow  \top\}_{1} \rangle$\\
   	Unsocial, Realistic		& $\langle \{a \leftarrow  \top\}_{0},\{b \leftarrow  \top\}_{1} \rangle$   & $\langle \{b \leftarrow  \top\}_{0}, \{b = \top \rightarrow ?\}_{1}, \{a \leftarrow  \top\}_{1} \rangle$\\
    Social, Optimistic		& $\langle \{b \leftarrow  \top\}_{0},\{a \leftarrow  \top\}_{2} \rangle$   & $\langle \{b \leftarrow  \bot\}_{1},\{b \leftarrow  \top\}_{3} \rangle$\\
    Social, Realistic		& $\langle \{b \leftarrow  \top\}_{0},\{b = \bot \rightarrow ?\}_{3:k-1}, \{a \leftarrow  \top\}_{k},\{b \leftarrow  \top\}_{k+1} \rangle$   & $\langle \{b = \top \rightarrow ?\}_{1},\{b \leftarrow  \bot\}_{2}\rangle$\\
 \end{tabular}
\caption{Agent's strategies for the collaborative Iterated Boolean Game. Message sequence notation simplified, as sending agent and temporal frame of reference obvious from location in table. Table depicts only the strategy for a single successful satisfaction of $\phi$ and $\psi$, omitting the return to the original state, or aborting in case of failed strategy detection.}
\label{tab:boolean_game_colab}
\end{table}

\begin{table}[h!]
\centering
\begin{tabular}{l | c|  c }
Modes  & $\phi$ & $\psi$ \\
\hline
    Unsocial, Optimistic    & 12.12 & 8\\
    Unsocial, Realistic   & 9.5& 9.5\\
    Social, Optimistic    & 16 & 12.12\\
    Social, Realistic   & 4.6 & 2.3\\
 \end{tabular}
\caption{Average number of time steps required per successful satisfaction of agents' formulae in function of strategy.}
\label{tab:boolean_game_colab_results}
\end{table}

This example is sufficiently simple that we can merely list all permutations of actions' effects and analyze the predicted outcome. \textbf{Unsocial, optimistic agents}: given $P(\mathbf{m}_n)$, $\phi$ is satisfied on 33\% of runs, $\psi$ is satisfied on 50\% of runs, and 17\% of runs satisfy neither formula, with an average run length of 4 time steps. In this case, $\psi$ is dominant since it allows for arbitrary time between transition from the base state until its completion, while $\phi$ requires an immediate transition from its second state to its third state (Equation \ref{eq:phi_psi}). \textbf{Unsocial but realistic agents}: given $P(\mathbf{m}_n)$, $\phi$ is satisfied on 40\% of runs, $\psi$ is satisfied on 40\% of runs, and 10\% of runs satisfy neither formula, with an average run length of 3.8 time steps. Compared to the prior case,  $\|P(\mathbf{m}_n)\|$ is smaller since $c_1$ can abort unsuccessful strategies, thus the revised success rates.  \textbf{Social but optimistic agents}: $\phi$ is uniquely satisfied on 25\% of runs, $\psi$ is uniquely satisfied on 33\% of runs, both are satisfied on 17\% of runs, and 25\% of runs satisfy neither formula, with an average run length of 4 time steps.  \textbf{Social and realistic agents}: $\phi$ is satisfied on 50\% of runs, whilst $\psi$ is satisfied 100\% of runs, with an average run length of 2.3 time steps. Aggregate results, showing average number of time steps expected for satisfaction of formulae are depicted in Table \ref{tab:boolean_game_colab_results}.

\begin{figure*}
\subfloat[]{\includegraphics[width = 0.4\columnwidth]{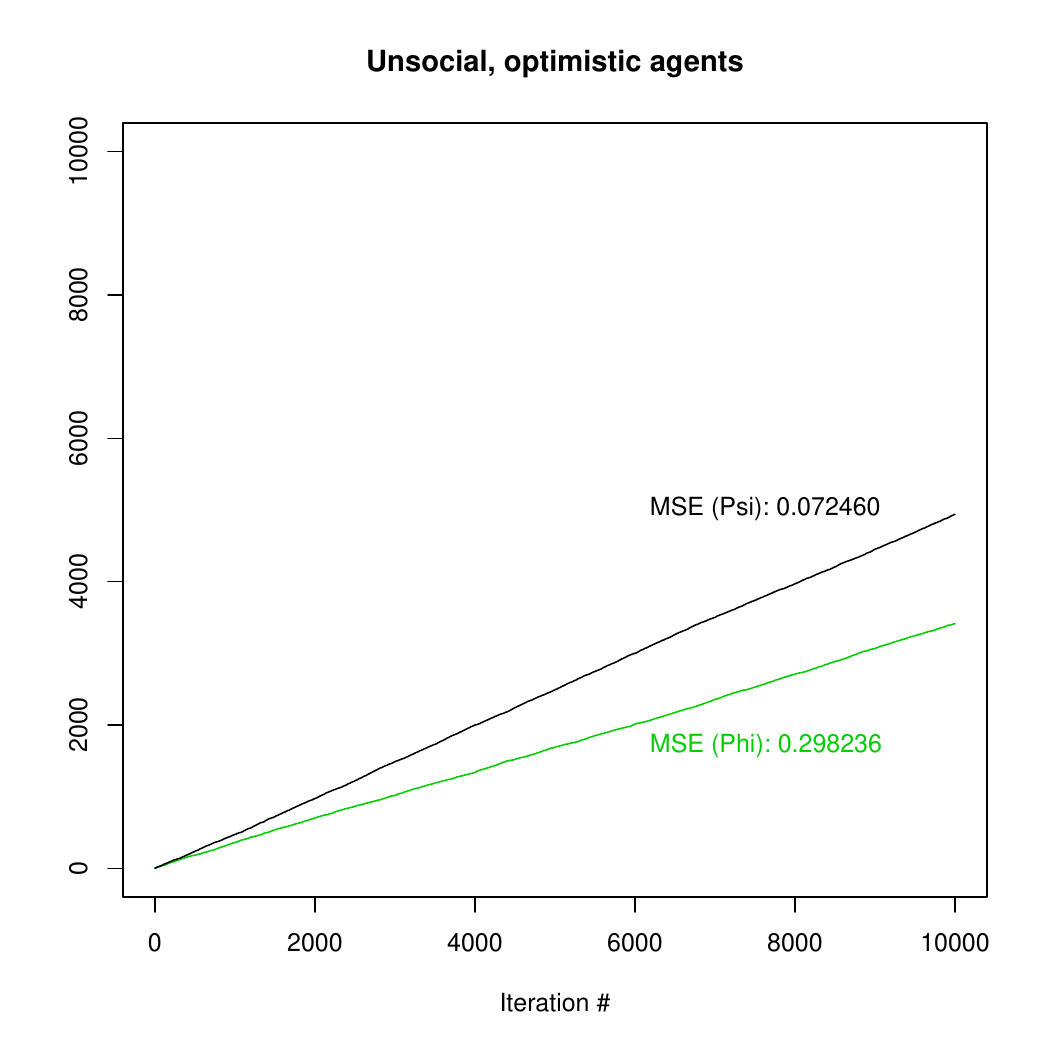}}
\subfloat[]{\includegraphics[width = 0.4\columnwidth]{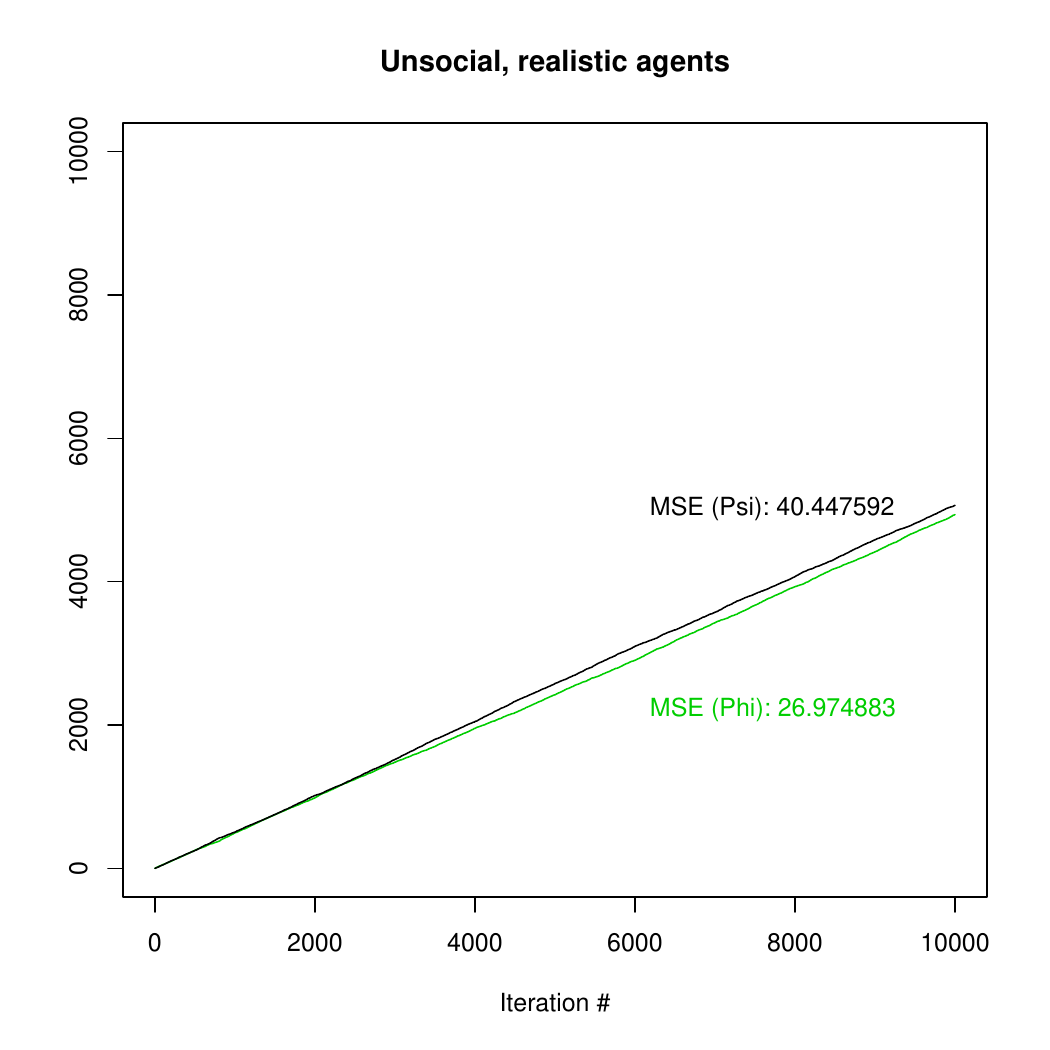}}\\
\subfloat[]{\includegraphics[width = 0.4\columnwidth]{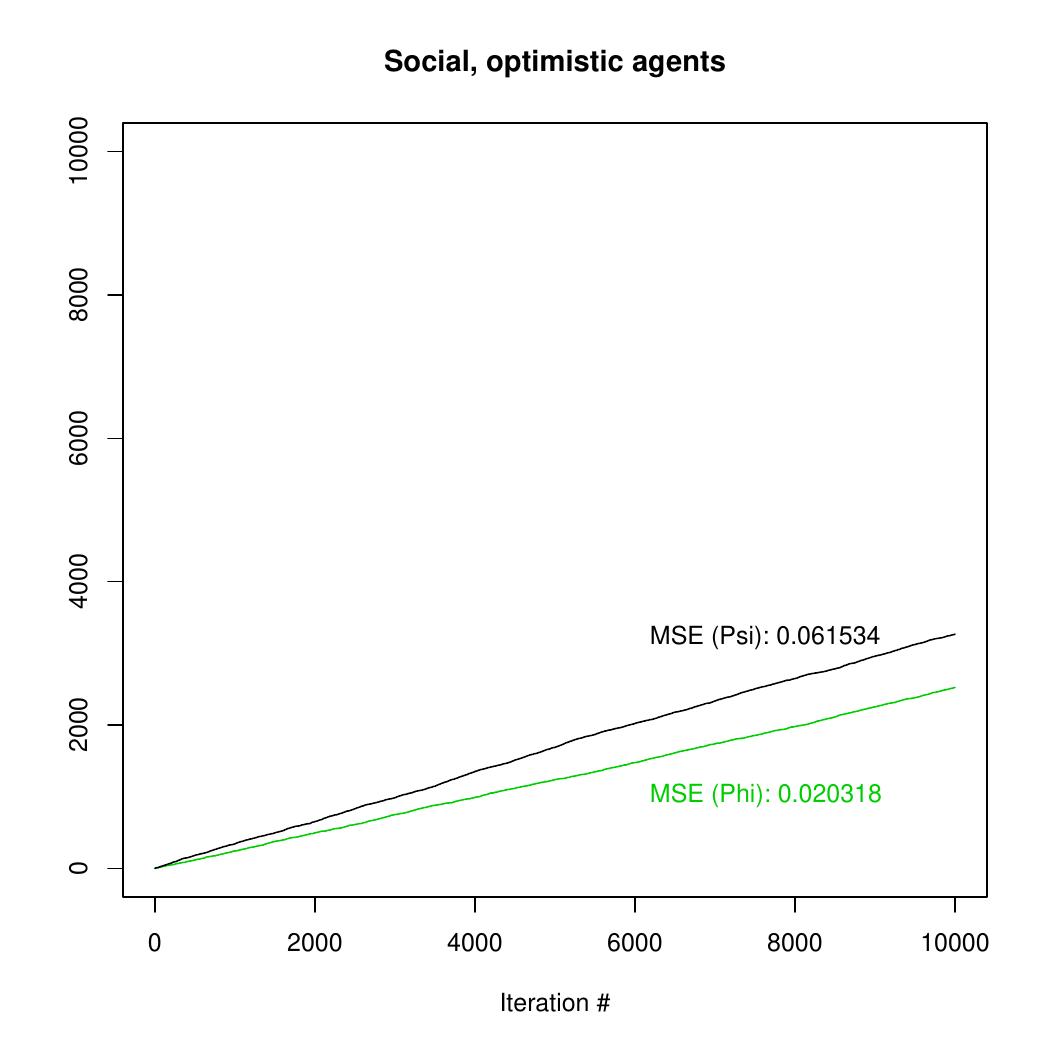}} 
\subfloat[]{\includegraphics[width = 0.4\columnwidth]{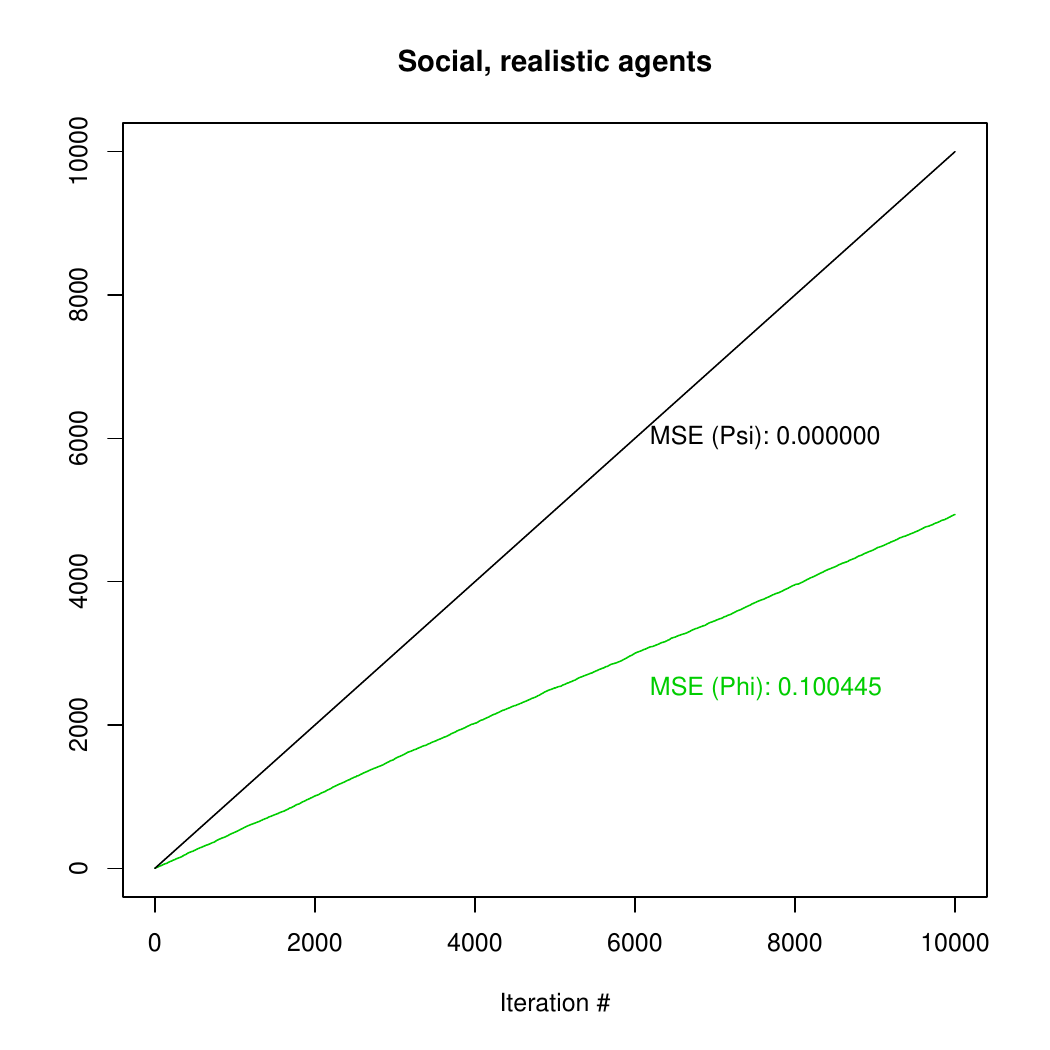}} 
\caption{Simulation results for the collaborative Iterated Boolean Game, showing formulae satisfaction per number of iterations.}
\label{fig:results_collab}
\end{figure*}

Regardless of social behavior, both success rates are improved if agents are aware of CPSs' temporal relativity, and adjust their behavior accordingly. Realistic behavior and collaboration result in the best possible result for both agents.Most interestingly, collaboration when ignoring
the temporal relativity of CPSs results in the worst possible outcome: overhead caused by naive collaboration acts to the detriment of both agents. Our simulation model yields the results displayed on Fig. \ref{fig:results_collab}, with Mean Square Error (MSE) in relation to predicted outcomes displayed.

\subsection{Evaluating the Adversarial Game}

\par Agent $c_0$'s goal is to satisfy a formula $\phi$, and agent $c_1$'s goal is to satisfy a formula $\psi$, given by:

\begin{equation}\label{eq:phi_psi}
\begin{aligned}
&	\phi : (\{\overline{a}\overline{b}\overline{c}\} \rightarrow \mathbf{X} \{\overline{a}\overline{b}c\} \rightarrow \mathbf{X} \{a\overline{b}c\}) \cup (\{\overline{a}b\overline{c}\} \rightarrow \mathbf{X} \{\overline{a}bc\} \rightarrow \mathbf{X} \{a\overline{b}c\}) \cup (\{ab\overline{c}\} \rightarrow \mathbf{X} \{abc\} \rightarrow \mathbf{X} \{\overline{a}bc\})\\
&	\psi : (\{\overline{a}b\overline{c}\} \rightarrow \mathbf{X} \{\overline{a}\overline{b}\overline{c}\} \cup  (\{\overline{a}bc\} \rightarrow \mathbf{X} \{\overline{a}\overline{b}c\}
\end{aligned}
\end{equation}

\begin{figure}[tbh]
\includegraphics[width=0.5\columnwidth]{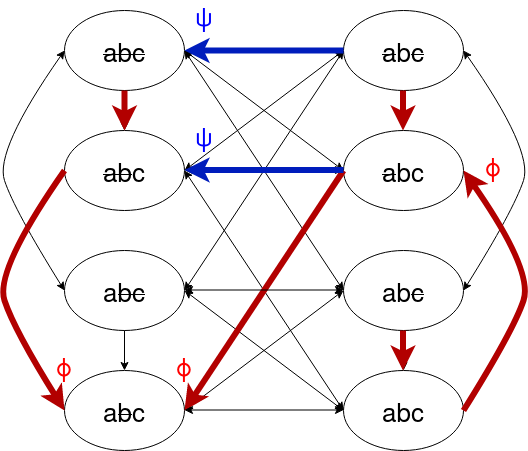}
\centering
\caption{Adversarial Iterated Boolean Game graph. All state transitions are bi-directional. Payoffs for $\phi$ and $\psi$ satisfaction directed runs highlighted and super-imposed.}
\label{fig:game}
\end{figure}

In this scenario, there are more options for $\phi$ satisfiability than for $\psi$; $\phi$ can be satisfied by three different sub-runs, but all require two successive successful state transitions. $\psi$ can only be satisfied by two different sub-runs, but each just requires one successful state transition.
Figure \ref{fig:game} depicts the specific game graph utilized, and Table \ref{tab:boolean_game_adver} depicts agents' strategies in function of their modes.

\begin{table}[h!]
\centering
\begin{tabular}{l | c|  c }
Modes  & $c_0$ & $c_1$ \\
\hline
    Unsocial, Optimistic    & $
    							\begin{cases}
    								\langle \{c \leftarrow \top \} \rangle, \{abc\} = \{\bot,\bot,\bot\}\\
    								\langle \{a \leftarrow \top\} \rangle, \{abc\} = \{\bot,\bot,\top\}\\
    								\langle \{c \leftarrow \top \} \rangle, \{abc\} = \{\bot,\top,\bot\}\\
    								\langle \{a \leftarrow \top \} \rangle, \{abc\} = \{\bot,\top,\top\}\\
    								\langle \{c \leftarrow \top \} \rangle, \{abc\} = \{\top,\bot,\bot\}\\
    								\langle \{c \leftarrow \bot \} \rangle, \{abc\} = \{\top,\bot,\top\}\\
    								\langle \{c \leftarrow \top \} \rangle, \{abc\} = \{\top,\top,\bot\}\\
    								\langle \{a \leftarrow \bot \} \rangle, \{abc\} = \{\top,\top,\top\}\\
    							\end{cases}$
    						& $
    							\begin{cases}
    								\langle \{b \leftarrow  \top\} \rangle, \{abc\} = \{\bot,\bot,*\}\\
    								\langle \{b \leftarrow  \bot\} \rangle, \{abc\} = \{\bot,\top,*\}\\
    								\langle \{b \leftarrow  \top\} \rangle, \{abc\} = \{\top,*,*\}
    							\end{cases}$\\
   	Unsocial, Realistic		& $
    							\begin{cases}
    								\langle \{c \leftarrow \top \},\{c = \top \rightarrow  ?\} \rangle, \{abc\} = \{\bot,\bot,\bot\}\\
    								\langle \{a \leftarrow \top\},\{a = \top \rightarrow  ?\} \rangle, \{abc\} = \{\bot,\bot,\top\}\\
    								\langle \{c \leftarrow \top \},\{c = \top \rightarrow  ?\} \rangle, \{abc\} = \{\bot,\top,\bot\}\\
    								\langle \{a \leftarrow \top \},\{a = \top \rightarrow  ?\} \rangle, \{abc\} = \{\bot,\top,\top\}\\
    								\langle \{c \leftarrow \top \},\{c = \top \rightarrow  ?\} \rangle, \{abc\} = \{\top,\bot,\bot\}\\
    								\langle \{c \leftarrow \bot \},\{c = \bot \rightarrow  ?\} \rangle, \{abc\} = \{\top,\bot,\top\}\\
    								\langle \{c \leftarrow \top \},\{c = \top \rightarrow  ?\} \rangle, \{abc\} = \{\top,\top,\bot\}\\
    								\langle \{a \leftarrow \bot \},\{a = \bot \rightarrow  ?\} \rangle, \{abc\} = \{\top,\top,\top\}\\
    							\end{cases}$
    						& $
    							\begin{cases}
    								\langle \{b \leftarrow  \top\},\{b = \top \rightarrow  ?\} \rangle, \{abc\} = \{\bot,\bot,*\}\\
    								\langle \{b \leftarrow  \bot\},\{b = \bot \rightarrow  ?\} \rangle, \{abc\} = \{\bot,\top,*\}\\
    								\langle \{b \leftarrow  \top\},\{b = \top \rightarrow  ?\} \rangle, \{abc\} = \{\top,*,*\}
    							\end{cases}$\\
    Social, Optimistic		& $
    							\begin{cases}
    								\langle \{\} \rangle, \{abc\} = \{\bot,\bot,\bot\}\\
    								\langle \{c \leftarrow \bot\} \rangle, \{abc\} = \{\bot,\bot,\top\}\\
    								\langle \{c \leftarrow \top\} \rangle, \{abc\} = \{\bot,\top,\bot\}\\
    								\langle \{a \leftarrow \top\} \rangle, \{abc\} = \{\bot,\top,\top\}\\
    								\langle \{a \leftarrow \bot\} \rangle, \{abc\} = \{\top,\bot,\bot\}\\
    								\langle \{c \leftarrow \bot\} \rangle, \{abc\} = \{\top,\bot,\top\}\\
    								\langle \{a \leftarrow \bot\} \rangle, \{abc\} = \{\top,\top,\bot\}\\
    								\langle \{c \leftarrow \bot\} \rangle, \{abc\} = \{\top,\top,\top\}\\
    							\end{cases}$
    						& $
    							\begin{cases}
    								\langle \{b \leftarrow \top\} \rangle, \{abc\} = \{\bot,\bot,\bot\}\\
    								\langle \{b \leftarrow \top\} \rangle, \{abc\} = \{\bot,\bot,\top\}\\
    								\langle \{b \leftarrow \bot\} \rangle, \{abc\} = \{\bot,\top,\bot\}\\
    								\langle \{\} \rangle, \{abc\} = \{\bot,\top,\top\}\\
    								\langle \{b \leftarrow \top\} \rangle, \{abc\} = \{\top,\bot,\bot\}\\
    								\langle \{b \leftarrow \top\} \rangle, \{abc\} = \{\top,\bot,\top\}\\
    								\langle \{\} \rangle, \{abc\} = \{\top,\top,\bot\}\\
    								\langle \{b \leftarrow \bot\} \rangle, \{abc\} = \{\top,\top,\top\}\\
    							\end{cases}$\\
    Social, Realistic		& $
    							\begin{cases}
    								\langle \{c \leftarrow \top\} \rangle, \{abc\} = \{\bot,\bot,\bot\}\\
    								\langle \{a \leftarrow \top\} \rangle, \{abc\} = \{\bot,\bot,\top\}\\
    								\langle \{a \leftarrow \top\} \rangle, \{abc\} = \{\bot,\top,\bot\}\\
    								\langle \{c \leftarrow \bot\} \rangle, \{abc\} = \{\bot,\top,\top\}\\
    								\langle \{a \leftarrow \bot\} \rangle, \{abc\} = \{\top,\bot,\bot\}\\
    								\langle \{c \leftarrow \bot\} \rangle, \{abc\} = \{\top,\bot,\top\}\\
    								\langle \{c \leftarrow \top\} \rangle, \{abc\} = \{\top,\top,\bot\}\\
    								\langle \{a \leftarrow \bot\} \rangle, \{abc\} = \{\top,\top,\top\}\\
    							\end{cases}$
    						& $
    							\begin{cases}
    								\langle \{b \leftarrow \top\} \rangle, \{abc\} = \{\bot,\bot,\bot\}\\
    								\langle \{b \leftarrow \top\} \rangle, \{abc\} = \{\bot,\bot,\top\}\\
    								\langle \{b \leftarrow \bot\} \rangle, \{abc\} = \{\bot,\top,\bot\}\\
    								\langle \{b \leftarrow \bot\} \rangle, \{abc\} = \{\bot,\top,\top\}\\
    								\langle \{b \leftarrow \top\} \rangle, \{abc\} = \{\top,\bot,\bot\}\\
    								\langle \{b \leftarrow \top\} \rangle, \{abc\} = \{\top,\bot,\top\}\\
    								\langle \{\} \rangle, \{abc\} = \{\top,\top,\bot\}\\
    								\langle \{\} \rangle, \{abc\} = \{\top,\top,\top\}\\
    							\end{cases}$\\
 \end{tabular}
\caption{Agent's strategies for the adversarial Iterated Boolean Game. Decision making is analyzed in function of current state for the adversarial version. }
\label{tab:boolean_game_adver}
\end{table}

For each specific pair of agent strategies (Table \ref{tab:boolean_game_adver}), we can construct the state sequence matrix $\mathbf{T}$, where row index $i$ denotes the state at time $k$ and column index $j$ represents the next state at time $k+1$, and entry $t_{i,j}$ denotes the probability of transitioning to state $j$ given current state $i$. The value for each entry  $t_{i,j}$ is given by Equation \ref{eq:probabilities}. For the Iterated Boolean Game analyzed here we obtain the following state sequence matrices (shortened for illustrative purposes; full matrices in Appendix). The set of most likely paths per PFA can be determined, for each matrix, through Algorithm \ref{alg:path}, for specific values of $\mathbb{P}(c_0,c_1)$, $\mathbb{P}(c_1,c_0)$, and $\mathbb{P}(c_0\|c_1)$. In this case, we arbitrarily set a fixed value of $\mathbb{P}(c_0,c_1) = 0.25$, $\mathbb{P}(c_1,c_0) = 0.25$, and $\mathbb{P}(c_0\|c_1) = 0.5$ (in practice, of course, these would be modeled as a random variable sampled from an appropriate distribution obtained from system profiling, but here simplicity suffices to illustrate our points).

\begin{equation}
\begin{aligned}
&    \mathbf{T}_{\text{Unsocial, Optimistic}} =
\begin{bmatrix}
    				0 & \frac{\mathbb{P}(c_0,c_1) + \mathbb{P}(c_1,c_0)}{g_0} & \frac{\mathbb{P}(c_1,c_0)}{g_0} & \hdots \\
    				0 & 0 & 0 & \hdots\\
    				\frac{\mathbb{P}(c_1,c_0)}{g_2}& \frac{\mathbb{P}(c_0\|c_1)}{g_2}&0 &\hdots \\
    				\vdots& \vdots&\vdots &\ddots \\
                 \end{bmatrix} = \begin{bmatrix}
		0		&0.4	&0.2	&0.4	&0	&0	&0	&0\\
	0		&0	&0	&0.333	&0	&0.167	&0.167	&0.333\\
	0.2	&0.4 &0	&0.4	&0	&0	&0	&0\\
	0		&0.2	&0.2	&0	&0	&0.4	&0	&0.2\\
	0		&0	&0	&0	&0	&0.2	&0.4	&0.4\\
	0		&0	&0	&0	&0.167	&0	&0.333	&0.5\\
	0		&0	&0	&0	&0	&0	&0	&1\\
	0		&0	&0	&0.5	&0.167	&0.167	&0.167	&0
	\end{bmatrix}\\
&   = \mathbf{T}_{\text{Unsocial, Realistic}}
\end{aligned}
\end{equation}

\begin{equation}
\begin{aligned}
&    \mathbf{T}_{\text{Social, Optimistic}} =
 \begin{bmatrix}
    				0  & \frac{\mathbb{P}(c_0,c_1)+2\mathbb{P}(c_1,c_0)+\mathbb{P}(c_0\|c_1)}{g_0}  & \frac{2\mathbb{P}(c_0,c_1)}{g_0} &\hdots\\
    				\frac{\mathbb{P}(c_0,c_1)}{g_1}   & 0 & \frac{\mathbb{P}(c_0\|c_1)}{g_1} &\hdots\\
    				 0   & \frac{\mathbb{P}(c_0\|c_1)}{g_2}  & 0 &\hdots\\
    				\vdots&\vdots&\vdots&\ddots\\
                 \end{bmatrix}=
                 \begin{bmatrix}
    0	& 0.714& 0.286& 0& 0&0 &0 &0 \\
    0.25 & 0& 0.5& 0.25& 0&0 & 0& 0\\
    0 & 0.667& 0& 0.333& 0& 0&0 &0 \\
    0.143 & 0.286& 0& 0& 0& 0& 0& 0.571\\
    0.2 &0 & 0.6& 0& 0& 0& 0.2&0 \\
    0 & 0& 0&0.2 & 0.2& 0& 0.4& 0.2\\
    0 & 0& 0.8& 0& 0.2& 0& 0&0 \\
    0 & 0& 0& 0&0.4 &0.2 &0.4 &0 \\
	\end{bmatrix}
\end{aligned}
\end{equation}

\begin{equation}
\begin{aligned}
&    \mathbf{T}_{\text{Social, Realistic}} =
 \begin{bmatrix}
    				0&\frac{\mathbb{P}(c_0,c_1)}{g_0}&\frac{\mathbb{P}(c_1,c_0)}{g_0}&\hdots\\
    				\frac{\mathbb{P}(c_1,c_0)}{g_1}&0&0&\hdots\\
    				\frac{\mathbb{P}(c_0,c_1)+\mathbb{P}(c_1,c_0)}{g_2}&0&0&\hdots\\
    				\vdots&\vdots&\vdots&\ddots\\
                 \end{bmatrix}=
                 \begin{bmatrix}
    	0	&0.167	&0.167	&0.5	&0.167	&0	&0	&0\\
	0.167	&0	&0	&0.333	&0	&0.167	&0	&0.333\\
	0.333	&0	&0	&0.167	&0.333	&0	&0.167	&0\\
	0.4	&0.2	&0.2	&0	&0	&0	&0	&0.2\\
	0.2	&0	&0.4	&0	&0	&0	&0.4	&0\\
	0	&0	&0	&0	&0.2	&0	&0.4	&0.4\\
	0	&0	&0.286	&0	&0.143	&0	&0	&0.571\\
	0	&0	&0	&0.8	&0	&0	&0.2	&0\\
	\end{bmatrix}
\end{aligned}
\end{equation}

This results in the following most likely state sequences for each configuration:

\begin{table}[h!]
\centering
\begin{tabular}{l | c}
Modes  & State Sequence \\
\hline
    Unsocial, Optimistic    & $ \langle \{\overline{a}bc\} ,\{a\overline{b}c\}, \{abc\} \rangle$\\
   	Unsocial, Realistic		& $ \langle \{\overline{a}bc\} ,\{a\overline{b}c\}, \{abc\} \rangle$\\
    Social, Optimistic		& $ \langle \{\overline{a}b\overline{c}\},\{\overline{a}\overline{b}c\}   \rangle$\\
    Social, Realistic		& $ \langle \{\overline{a}\overline{b}\overline{c}\},\{\overline{a}bc\}  \rangle$\\

 \end{tabular}
\caption{Most likely state sequences in the Iterated Boolean Game. }
\label{tab:state_seq}
\end{table}

\begin{figure*}
\subfloat[]{\includegraphics[width = 0.45\columnwidth]{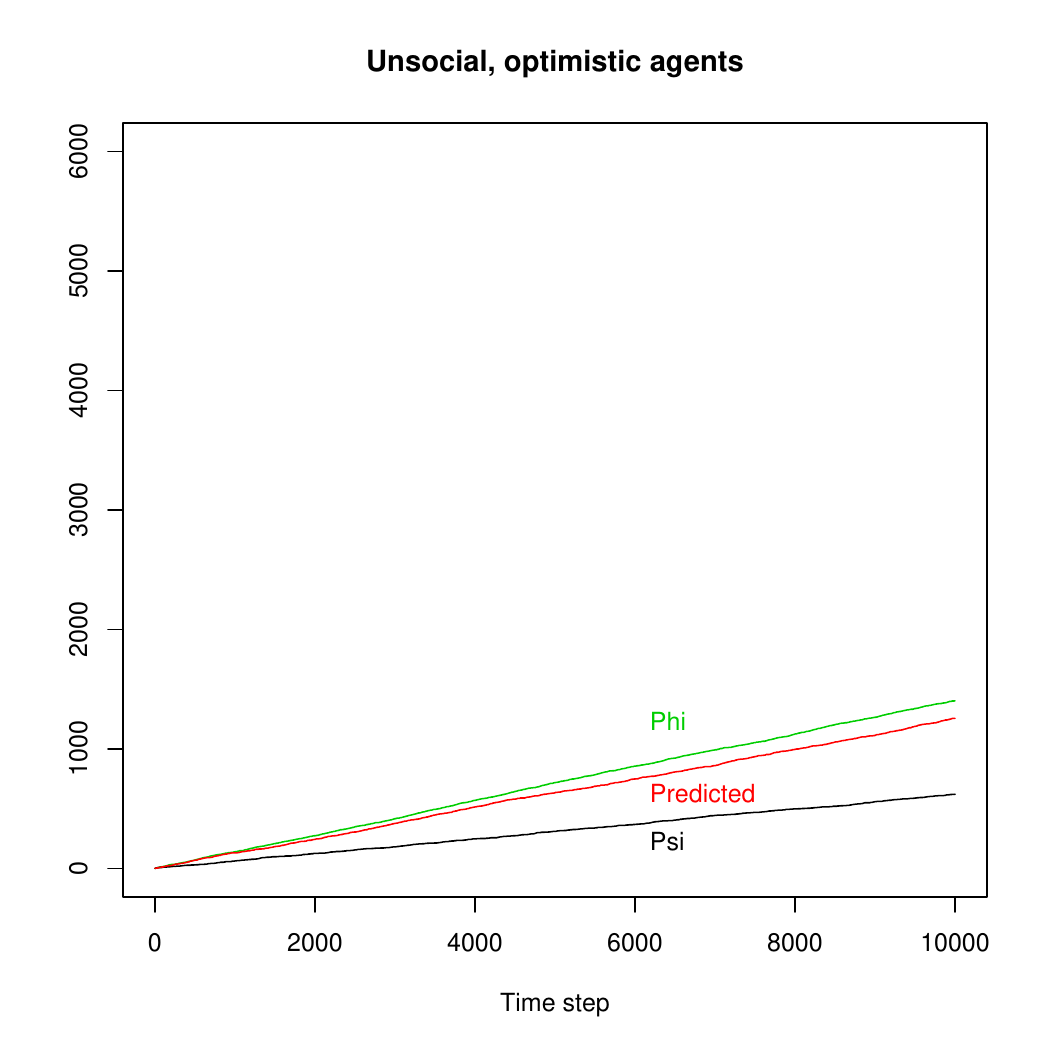}}
\subfloat[]{\includegraphics[width = 0.45\columnwidth]{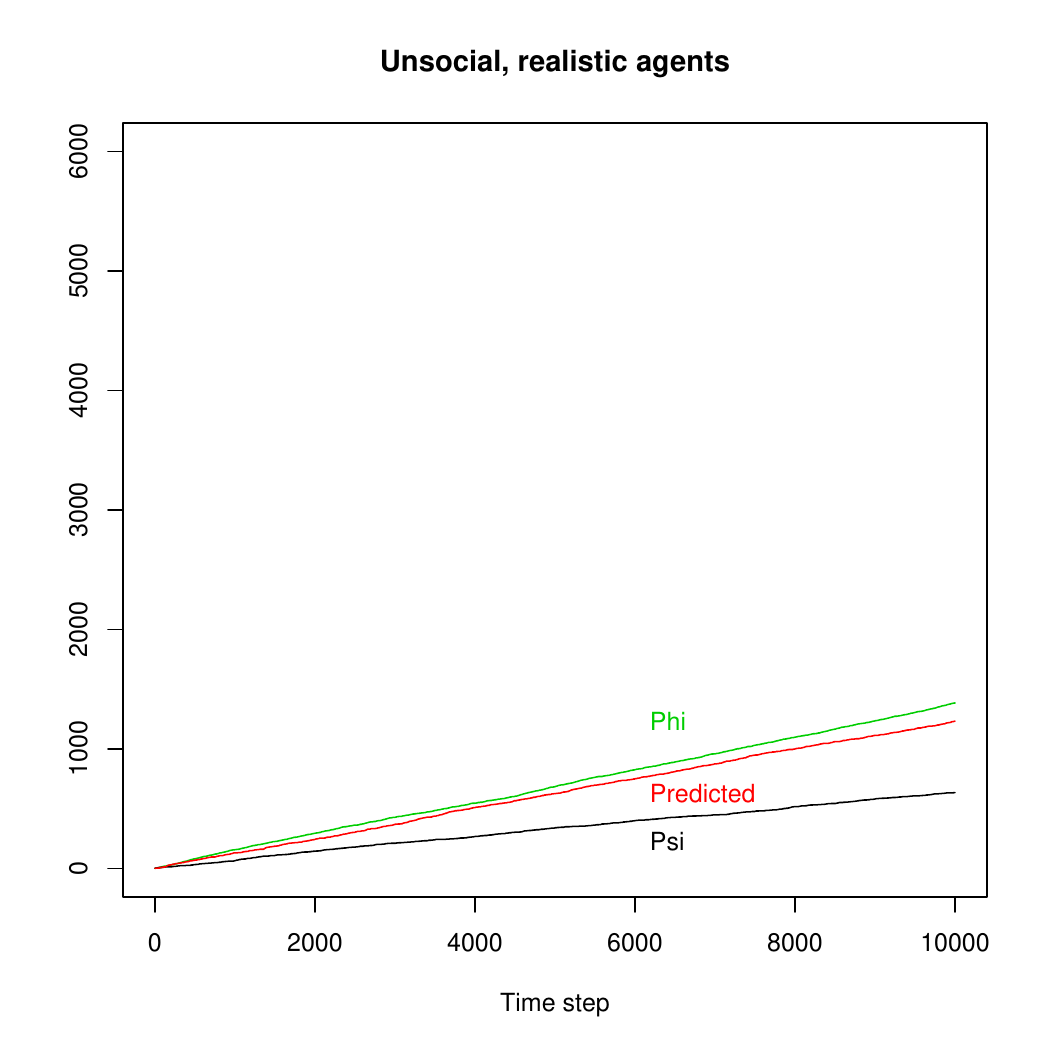}}\\
\subfloat[]{\includegraphics[width = 0.45\columnwidth]{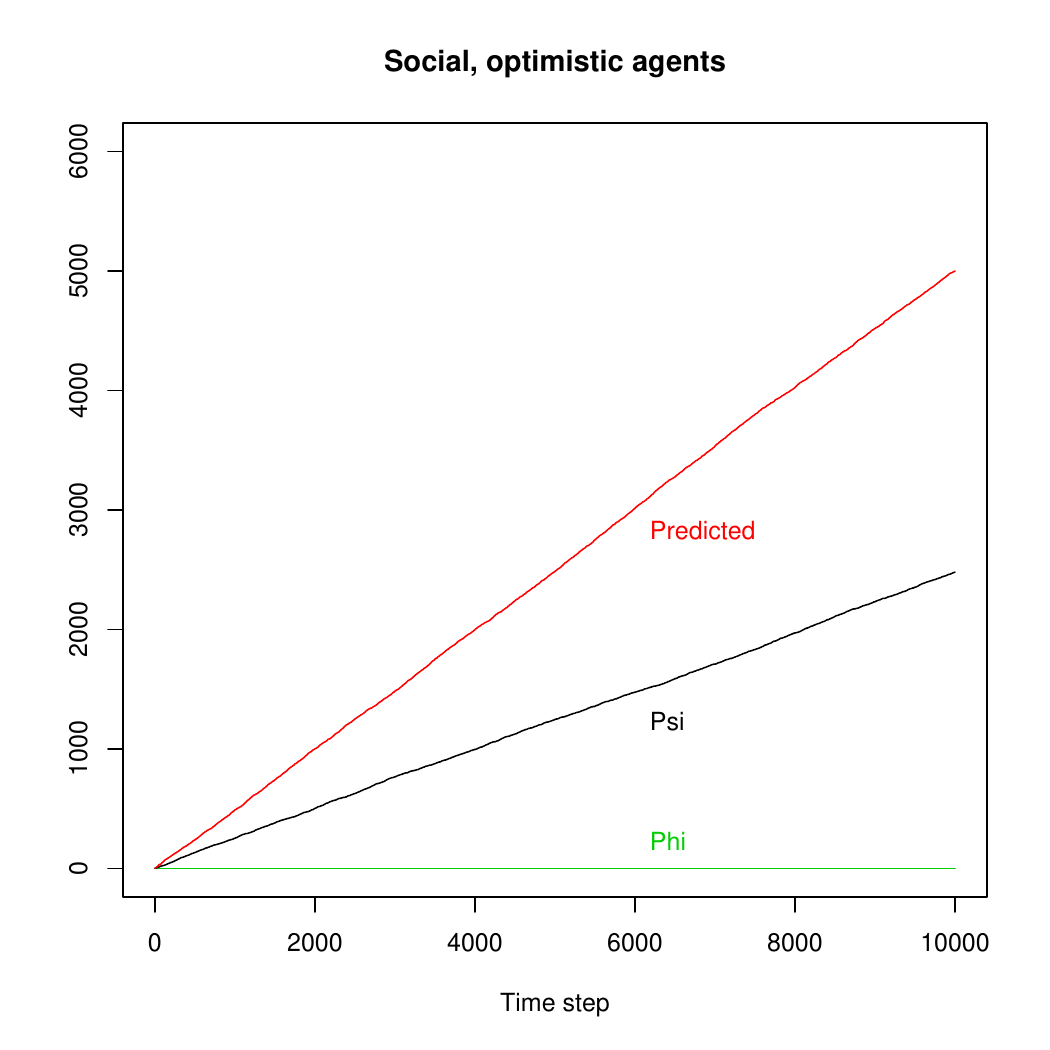}} 
\subfloat[]{\includegraphics[width = 0.45\columnwidth]{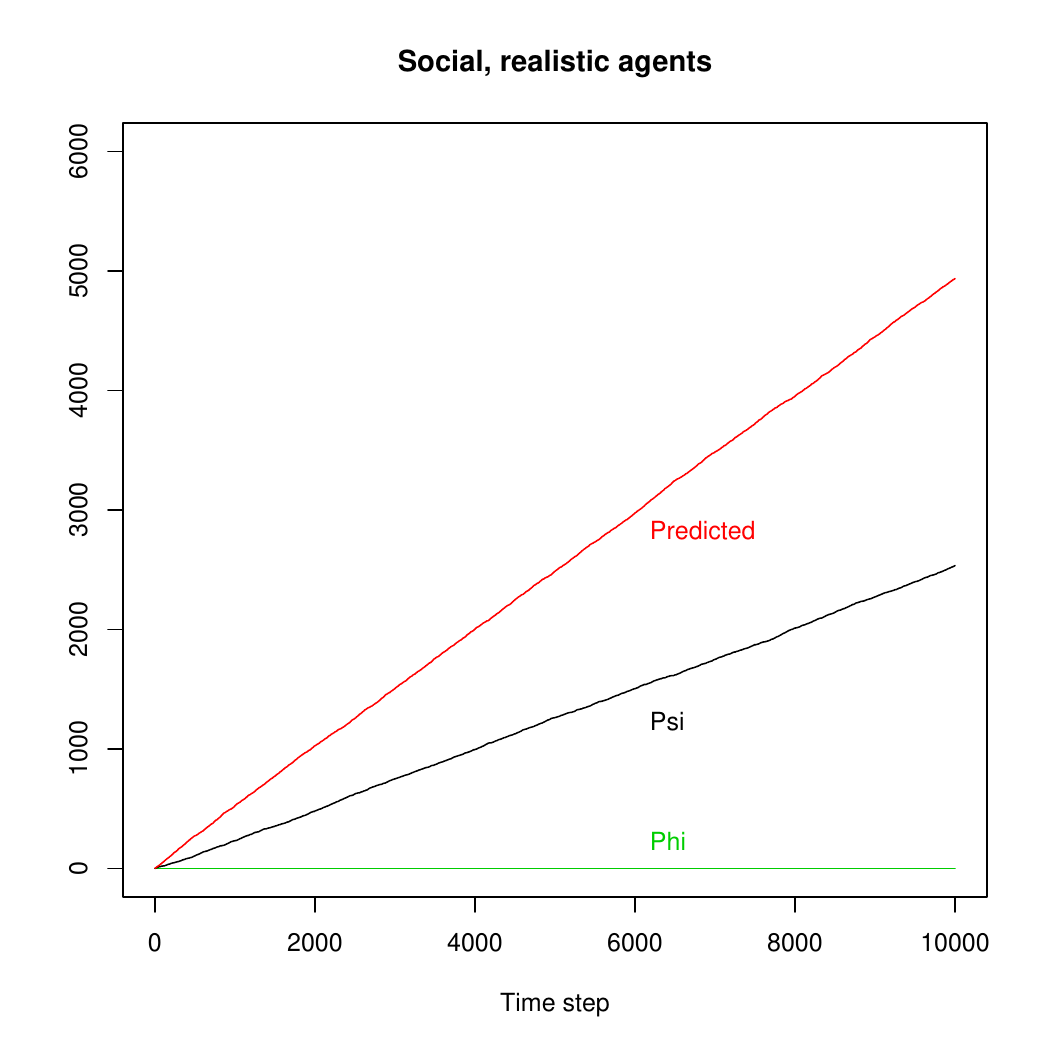}} 
\caption{Simulation results for the adversarial Iterated Boolean Game, showing formulae satisfaction and predicted state sequence occurrence per time step.}
\label{fig:strategies_adv}
\end{figure*}

Our simulation model yields the results displayed on Fig. \ref{fig:strategies_adv}.

\subsection{Discussion}

Simulation results for the collaborative game show are consistent with predicted values, with very small Mean Square Error (MSE) for all cases except Unsocial and Realistic agents, where our model is pessimistic: being aware of environment progression and aborting failed runs preemptively seems to work out better than expected (it's possible that this is an artifact of the small state space).  In the adversarial game, Unsocial agents lead to the satisfaction of one agent's formula, closely aligned with predicted most likely state sequence (in this case, we see $\Phi$ satisfaction because there is no active adversarial action, thus the higher number of options for $\Phi$ leads to victory). When agents behave socially, $\Psi$ allows for a dominant strategy, but still far below the most likely state path as predicted by our model. In all cases, it is beneficial for all system agents to be designed such that they are aware, and actively employ knowledge, of the non-deterministic properties of Cyber Physical Games. 
\par In cases where CPSs are designed in a bottom-up approach, scenarios will arise when an agent is added to a game where other agents have irrevocably committed to satisfying certain formulae. In these cases, the novel agent has an advantage if it has flexibility in choosing its formula: the described model can be used to analyze the system and extract the most likely path of states \textit{a priori}. Even in collaborative cases, this has relevance for system optimization.
\par Finally, we offer the conjecture that the "Multi-armed bandits" \cite{wang2008algorithms} algorithmic framework is the ideal way to formulate and guide the behavior of agents in Cyber Physical Games, as it provides mechanisms that can be applied to the described problem characteristics, and an extensive body of work has been developed to deal with different probability distributions for each action.

\section{Related Work}\label{sec:related}

Timing properties of Cyber Physical Systems and how they impact both design and operations of such systems have been extensively studied. Time-criticality in the presence of non-determinism is explored by Liu et al \cite{10.1145/3185502}, introducing "time-soundness"; similar properties are achieved by Martin et al \cite{10.1145/3134841} in the context of choreographing wireless devices' actions across both time and space using an Extended Kalman Filter to estimate all devices' positions and clock errors. Rosales and Paulitsch formalize these sort of techniques using a tagged-signal-model model of computation for time semantics \cite{10.1145/3386244}, whilst Khayatian et al \cite{10.1145/3516449} examine effects and prevention techniques for the graceful handling of uncontrollable timing errors (e.g., network failure), akin to the the study of non-determinism described in this paper.
\par Formal games as a model for the behavior of CPSs has been applied to both design and security. For example, Krichene et al \cite{10.1145/3078620} show how the routing game models congestion in different types of networks, specifically in cases where the joint decision of all players determines the costs of each path; they show that, under some constraints, the joint decision of the players remains within a small distance of the set of equilibria. Bakker et al \cite{10.1145/3384676} use hypergames, which form an extension of game theory that enables modeling strategic interactions where the players may have significantly different perceptions of the game(s) they are playing, to model large systems with continuous variables for the purposes of cyber-security, later extending this framework to 	develop hypergame-based defender strategies that are robust to deception and do not rely on attack detection \cite{10.1145/3439430}.
\par The use and emergence of artificial intelligence, at several levels, has been explored in the context of CPSs as self-aware systems; Bellman et al \cite{10.1145/3375716} show that these types of systems can perform decisions autonomously at runtime in a self-explanatory way, even in the presence of complex unforeseeable environmental dynamics (such as the ones described in this paper: we illustrated Cyber Physical Games with agents demonstrating a fairly low level of intelligence). This idea has been built upon by Esterle and Brown \cite{10.1145/3375403}, who extend the idea of the self-awareness of individual systems toward networked self-awareness.  The need to model these  emergent system properties has led to the development of  formal calculi for this purpose: e.g., Bakirtzis et al \cite{10.1145/3461669} developed a category-theoretic framework to make different types of composition explicit in the modeling and analysis of cyber-physical systems, making 	 system behavior hierarchically decomposed and related to a system architecture using the systems-as-algebras paradigm.
\par To the best of our knowledge, this paper in the first that formalizes non-determinism in CPSs as a game, and shows its equivalence to probabilistic finite automata. As Lee \cite{10.1145/2912149} has stated: "\textit{...incompleteness of determinism... has profound consequences.}".

\section{Conclusions}\label{sec:conclusions}

We presented a formal description of Cyber Physical Games, and how they arise from emergent properties of multi-agent Cyber Physical Systems. Critically, we showed that multiple agents operating within a Cyber Physical System give rise to behavior that can be modeled as, and is in fact equivalent to, probabilistic finite state automata. This behavior can hinder the execution of all agents in a Cyber Physical Game, if it is not taken into account; thus, we introduced an algorithm for determining the most likely sequence of events.
\par  IT is clear that, at least at a fundamental micro level, Cyber Physical Systems can never be free of non-determinism. It affects every interaction between cyber systems and the underlying environment, making the outcome of every action tuple probabilistic. What is not clear, though, is how impactful this is at the macro level; what are the implications for system performance, reliability, security, etc.? In other words, are Cyber Physical Games useful modeling constructs at the micro level that can be abstracted away in favor of deterministic constructs at the macro scale; or, must we model non-deterministically all the way up? If so, how to best engineer each individual agent, or groups of agents?
\par Answering these questions requires further research on the theoretical and empirical fronts. On the theoretical side, further development of the theory of Cyber Physical Games -particularly, with many agents- and study of the emergent statistical properties is required. Again, we believe the Multi-armed bandit framework is a promising direction, but many other approaches are possible. On the empirical front, profiling of instances of Cyber Physical Systems can tell us what are the relevant probability distributions, and how much, at the macro level, are systems afflicted by non-determinism.

\bibliography{sample}

\begin{thebibliography}{10}
\urlstyle{rm}
\expandafter\ifx\csname url\endcsname\relax
  \def\url#1{\texttt{#1}}\fi
\expandafter\ifx\csname urlprefix\endcsname\relax\def\urlprefix{URL }\fi
\expandafter\ifx\csname doiprefix\endcsname\relax\def\doiprefix{DOI: }\fi
\providecommand{\bibinfo}[2]{#2}
\providecommand{\eprint}[2][]{\url{#2}}

\bibitem{jazdi2014cyber}
\bibinfo{author}{Jazdi, N.}
\newblock \bibinfo{title}{Cyber physical systems in the context of industry
  4.0}.
\newblock In \emph{\bibinfo{booktitle}{2014 IEEE international conference on
  automation, quality and testing, robotics}}, \bibinfo{pages}{1--4}
  (\bibinfo{organization}{IEEE}, \bibinfo{year}{2014}).

\bibitem{xin2015cyber}
\bibinfo{author}{Xin, S.} \emph{et~al.}
\newblock \bibinfo{journal}{\bibinfo{title}{Cyber-physical modeling and
  cyber-contingency assessment of hierarchical control systems}}.
\newblock {\emph{\JournalTitle{IEEE Transactions on Smart Grid}}}
  \textbf{\bibinfo{volume}{6}}, \bibinfo{pages}{2375--2385}
  (\bibinfo{year}{2015}).

\bibitem{rose2015internet}
\bibinfo{author}{Rose, K.}, \bibinfo{author}{Eldridge, S.} \&
  \bibinfo{author}{Chapin, L.}
\newblock \bibinfo{journal}{\bibinfo{title}{The internet of things: An
  overview}}.
\newblock {\emph{\JournalTitle{The internet society (ISOC)}}}
  \textbf{\bibinfo{volume}{80}}, \bibinfo{pages}{1--53} (\bibinfo{year}{2015}).

\bibitem{singh2021emergence}
\bibinfo{author}{Singh, K.~K.}, \bibinfo{author}{Nayyar, A.},
  \bibinfo{author}{Tanwar, S.} \& \bibinfo{author}{Abouhawwash, M.}
\newblock \bibinfo{journal}{\bibinfo{title}{Emergence of cyber physical system
  and iot in smart automation and robotics}}.
\newblock {\emph{\JournalTitle{Computer Engineering in Automation}}}
  (\bibinfo{year}{2021}).

\bibitem{luckeneder2018systematic}
\bibinfo{author}{Luckeneder, C.} \& \bibinfo{author}{Kaindl, H.}
\newblock \bibinfo{title}{Systematic top-down design of cyber-physical models
  with integrated validation and formal verification}.
\newblock In \emph{\bibinfo{booktitle}{Proceedings of the 40th International
  Conference on Software Engineering: Companion Proceeedings}},
  \bibinfo{pages}{274--275} (\bibinfo{year}{2018}).

\bibitem{lee2006cyber}
\bibinfo{author}{Lee, E.~A.}
\newblock \bibinfo{title}{Cyber-physical systems-are computing foundations
  adequate}.
\newblock In \emph{\bibinfo{booktitle}{Position paper for NSF workshop on
  cyber-physical systems: research motivation, techniques and roadmap}},
  vol.~\bibinfo{volume}{2}, \bibinfo{pages}{1--9} (\bibinfo{year}{2006}).

\bibitem{leitao2016industrial}
\bibinfo{author}{Leit{\~a}o, P.}, \bibinfo{author}{Colombo, A.~W.} \&
  \bibinfo{author}{Karnouskos, S.}
\newblock \bibinfo{journal}{\bibinfo{title}{Industrial automation based on
  cyber-physical systems technologies: Prototype implementations and
  challenges}}.
\newblock {\emph{\JournalTitle{Computers in industry}}}
  \textbf{\bibinfo{volume}{81}}, \bibinfo{pages}{11--25}
  (\bibinfo{year}{2016}).

\bibitem{nanda2019internet}
\bibinfo{author}{Nanda, A.}, \bibinfo{author}{Puthal, D.},
  \bibinfo{author}{Rodrigues, J.~J.} \& \bibinfo{author}{Kozlov, S.~A.}
\newblock \bibinfo{journal}{\bibinfo{title}{Internet of autonomous vehicles
  communications security: overview, issues, and directions}}.
\newblock {\emph{\JournalTitle{IEEE Wireless Communications}}}
  \textbf{\bibinfo{volume}{26}}, \bibinfo{pages}{60--65}
  (\bibinfo{year}{2019}).

\bibitem{leitao2016smart}
\bibinfo{author}{Leitao, P.} \emph{et~al.}
\newblock \bibinfo{journal}{\bibinfo{title}{Smart agents in industrial
  cyber--physical systems}}.
\newblock {\emph{\JournalTitle{Proceedings of the IEEE}}}
  \textbf{\bibinfo{volume}{104}}, \bibinfo{pages}{1086--1101}
  (\bibinfo{year}{2016}).

\bibitem{GUTIERREZ2023103806}
\bibinfo{author}{Gutierrez, J.}, \bibinfo{author}{Kowara, S.},
  \bibinfo{author}{Kraus, S.}, \bibinfo{author}{Steeples, T.} \&
  \bibinfo{author}{Wooldridge, M.}
\newblock \bibinfo{journal}{\bibinfo{title}{Cooperative concurrent games}}.
\newblock {\emph{\JournalTitle{Artificial Intelligence}}}
  \textbf{\bibinfo{volume}{314}}, \bibinfo{pages}{103806},
  \doiprefix\url{https://doi.org/10.1016/j.artint.2022.103806}
  (\bibinfo{year}{2023}).

\bibitem{gutierrez2021equilibrium}
\bibinfo{author}{Gutierrez, J.}, \bibinfo{author}{Najib, M.},
  \bibinfo{author}{Perelli, G.} \& \bibinfo{author}{Wooldridge, M.}
\newblock \bibinfo{journal}{\bibinfo{title}{Equilibrium design for concurrent
  games}}.
\newblock {\emph{\JournalTitle{arXiv preprint arXiv:2106.10192}}}
  (\bibinfo{year}{2021}).

\bibitem{gutierrez2017reasoning}
\bibinfo{author}{Gutierrez, J.}, \bibinfo{author}{Harrenstein, P.} \&
  \bibinfo{author}{Wooldridge, M.}
\newblock \bibinfo{journal}{\bibinfo{title}{Reasoning about equilibria in
  game-like concurrent systems}}.
\newblock {\emph{\JournalTitle{Annals of Pure and Applied Logic}}}
  \textbf{\bibinfo{volume}{168}}, \bibinfo{pages}{373--403}
  (\bibinfo{year}{2017}).

\bibitem{lamport2019time}
\bibinfo{author}{Lamport, L.}
\newblock \bibinfo{title}{Time, clocks, and the ordering of events in a
  distributed system}.
\newblock In \emph{\bibinfo{booktitle}{Concurrency: the Works of Leslie
  Lamport}}, \bibinfo{pages}{179--196} (\bibinfo{year}{2019}).

\bibitem{russell2010artificial}
\bibinfo{author}{Russell, S.~J.} \& \bibinfo{author}{Norvig, P.}
\newblock \emph{\bibinfo{title}{Artificial intelligence a modern approach}}
  (\bibinfo{publisher}{London}, \bibinfo{year}{2010}).

\bibitem{10.1145/3047402}
\bibinfo{author}{Kuo, T.}
\newblock \bibinfo{journal}{\bibinfo{title}{Introduction}}.
\newblock {\emph{\JournalTitle{ACM Trans. Cyber-Phys. Syst.}}}
  \textbf{\bibinfo{volume}{1}}, \doiprefix\url{10.1145/3047402}
  (\bibinfo{year}{2017}).

\bibitem{10.1145/2912149}
\bibinfo{author}{Lee, E.~A.}
\newblock \bibinfo{journal}{\bibinfo{title}{Fundamental limits of
  cyber-physical systems modeling}}.
\newblock {\emph{\JournalTitle{ACM Trans. Cyber-Phys. Syst.}}}
  \textbf{\bibinfo{volume}{1}}, \doiprefix\url{10.1145/2912149}
  (\bibinfo{year}{2016}).

\bibitem{parsons2012game}
\bibinfo{author}{Parsons, S.~D.}, \bibinfo{author}{Gymtrasiewicz, P.} \&
  \bibinfo{author}{Wooldridge, M.}
\newblock \emph{\bibinfo{title}{Game theory and decision theory in agent-based
  systems}}, vol.~\bibinfo{volume}{5} (\bibinfo{publisher}{Springer Science \&
  Business Media}, \bibinfo{year}{2012}).

\bibitem{jantsch2003modeling}
\bibinfo{author}{Jantsch, A.}
\newblock \emph{\bibinfo{title}{Modeling Embedded Systems and SoC's:
  Concurrency and Time in Models of Computation}}
  (\bibinfo{publisher}{Elsevier}, \bibinfo{year}{2003}).

\bibitem{lamport2019part}
\bibinfo{author}{Lamport, L.}
\newblock \bibinfo{title}{The part-time parliament}.
\newblock In \emph{\bibinfo{booktitle}{Concurrency: the Works of Leslie
  Lamport}}, \bibinfo{pages}{277--317} (\bibinfo{year}{2019}).

\bibitem{gutierrez2015iterated}
\bibinfo{author}{Gutierrez, J.}, \bibinfo{author}{Harrenstein, P.} \&
  \bibinfo{author}{Wooldridge, M.}
\newblock \bibinfo{journal}{\bibinfo{title}{Iterated boolean games}}.
\newblock {\emph{\JournalTitle{Information and Computation}}}
  \textbf{\bibinfo{volume}{242}}, \bibinfo{pages}{53--79}
  (\bibinfo{year}{2015}).

\bibitem{bonzon2006boolean}
\bibinfo{author}{Bonzon, E.}, \bibinfo{author}{Lagasquie-Schiex, M.-C.},
  \bibinfo{author}{Lang, J.} \& \bibinfo{author}{Zanuttini, B.}
\newblock \bibinfo{title}{Boolean games revisited}.
\newblock In \emph{\bibinfo{booktitle}{ECAI}}, vol. \bibinfo{volume}{141},
  \bibinfo{pages}{265--269} (\bibinfo{year}{2006}).

\bibitem{rozier2011linear}
\bibinfo{author}{Rozier, K.~Y.}
\newblock \bibinfo{journal}{\bibinfo{title}{Linear temporal logic symbolic
  model checking}}.
\newblock {\emph{\JournalTitle{Computer Science Review}}}
  \textbf{\bibinfo{volume}{5}}, \bibinfo{pages}{163--203}
  (\bibinfo{year}{2011}).

\bibitem{wen1994folk}
\bibinfo{author}{Wen, Q.}
\newblock \bibinfo{journal}{\bibinfo{title}{The" folk theorem" for repeated
  games with complete information}}.
\newblock {\emph{\JournalTitle{Econometrica: Journal of the Econometric
  Society}}} \bibinfo{pages}{949--954} (\bibinfo{year}{1994}).

\bibitem{devroye1996random}
\bibinfo{author}{Devroye, L.} \& \bibinfo{author}{Kamoun, O.}
\newblock \bibinfo{title}{Random minimax game trees}.
\newblock In \emph{\bibinfo{booktitle}{Random Discrete Structures}},
  \bibinfo{pages}{55--80} (\bibinfo{organization}{Springer},
  \bibinfo{year}{1996}).

\bibitem{renault2006value}
\bibinfo{author}{Renault, J.}
\newblock \bibinfo{journal}{\bibinfo{title}{The value of markov chain games
  with lack of information on one side}}.
\newblock {\emph{\JournalTitle{Mathematics of Operations Research}}}
  \textbf{\bibinfo{volume}{31}}, \bibinfo{pages}{490--512}
  (\bibinfo{year}{2006}).

\bibitem{10.1093/logcom/exs049}
\bibinfo{author}{de~la Higuera, C.} \& \bibinfo{author}{Oncina, J.}
\newblock \bibinfo{journal}{\bibinfo{title}{{The most probable string: an
  algorithmic study}}}.
\newblock {\emph{\JournalTitle{Journal of Logic and Computation}}}
  \textbf{\bibinfo{volume}{24}}, \bibinfo{pages}{311--330},
  \doiprefix\url{10.1093/logcom/exs049} (\bibinfo{year}{2013}).
\newblock
  \eprint{https://academic.oup.com/logcom/article-pdf/24/2/311/3230951/exs049.pdf}.

\bibitem{kwak2017central}
\bibinfo{author}{Kwak, S.~G.} \& \bibinfo{author}{Kim, J.~H.}
\newblock \bibinfo{journal}{\bibinfo{title}{Central limit theorem: the
  cornerstone of modern statistics}}.
\newblock {\emph{\JournalTitle{Korean journal of anesthesiology}}}
  \textbf{\bibinfo{volume}{70}}, \bibinfo{pages}{144} (\bibinfo{year}{2017}).

\bibitem{vidal2005probabilistic}
\bibinfo{author}{Vidal, E.}, \bibinfo{author}{Thollard, F.},
  \bibinfo{author}{De~La~Higuera, C.}, \bibinfo{author}{Casacuberta, F.} \&
  \bibinfo{author}{Carrasco, R.~C.}
\newblock \bibinfo{journal}{\bibinfo{title}{Probabilistic finite-state
  machines-part i}}.
\newblock {\emph{\JournalTitle{IEEE transactions on pattern analysis and
  machine intelligence}}} \textbf{\bibinfo{volume}{27}},
  \bibinfo{pages}{1013--1025} (\bibinfo{year}{2005}).

\bibitem{buttelmann1971generalized}
\bibinfo{author}{Buttelmann, H.}
\newblock \bibinfo{title}{On generalized finite automata and unrestricted
  generative grammars}.
\newblock In \emph{\bibinfo{booktitle}{Proceedings of the third annual ACM
  symposium on Theory of computing}}, \bibinfo{pages}{63--77}
  (\bibinfo{year}{1971}).

\bibitem{cognetta2018incremental}
\bibinfo{author}{Cognetta, M.}, \bibinfo{author}{Han, Y.-S.} \&
  \bibinfo{author}{Kwon, S.~C.}
\newblock \bibinfo{title}{Incremental computation of infix probabilities for
  probabilistic finite automata}.
\newblock In \emph{\bibinfo{booktitle}{Proceedings of the 2018 Conference on
  Empirical Methods in Natural Language Processing}},
  \bibinfo{pages}{2732--2741} (\bibinfo{year}{2018}).

\bibitem{chatterjee2010probabilistic}
\bibinfo{author}{Chatterjee, K.} \& \bibinfo{author}{Henzinger, T.~A.}
\newblock \bibinfo{title}{Probabilistic automata on infinite words:
  Decidability and undecidability results}.
\newblock In \emph{\bibinfo{booktitle}{International Symposium on Automated
  Technology for Verification and Analysis}}, \bibinfo{pages}{1--16}
  (\bibinfo{organization}{Springer}, \bibinfo{year}{2010}).

\bibitem{de2013computing}
\bibinfo{author}{De~la Higuera, C.} \& \bibinfo{author}{Oncina, J.}
\newblock \bibinfo{title}{Computing the most probable string with a
  probabilistic finite state machine}.
\newblock In \emph{\bibinfo{booktitle}{Proceedings of the 11th International
  Conference on Finite State Methods and Natural Language Processing}},
  \bibinfo{pages}{1--8} (\bibinfo{year}{2013}).

\bibitem{durstenfeld1964algorithm}
\bibinfo{author}{Durstenfeld, R.}
\newblock \bibinfo{journal}{\bibinfo{title}{Algorithm 235: random
  permutation}}.
\newblock {\emph{\JournalTitle{Communications of the ACM}}}
  \textbf{\bibinfo{volume}{7}}, \bibinfo{pages}{420} (\bibinfo{year}{1964}).

\bibitem{wang2008algorithms}
\bibinfo{author}{Wang, Y.}, \bibinfo{author}{Audibert, J.-Y.} \&
  \bibinfo{author}{Munos, R.}
\newblock \bibinfo{journal}{\bibinfo{title}{Algorithms for infinitely
  many-armed bandits}}.
\newblock {\emph{\JournalTitle{Advances in Neural Information Processing
  Systems}}} \textbf{\bibinfo{volume}{21}} (\bibinfo{year}{2008}).

\bibitem{10.1145/3185502}
\bibinfo{author}{Liu, G.}, \bibinfo{author}{Jiang, C.} \&
  \bibinfo{author}{Zhou, M.}
\newblock \bibinfo{journal}{\bibinfo{title}{Time-soundness of time petri nets
  modelling time-critical systems}}.
\newblock {\emph{\JournalTitle{ACM Trans. Cyber-Phys. Syst.}}}
  \textbf{\bibinfo{volume}{2}}, \doiprefix\url{10.1145/3185502}
  (\bibinfo{year}{2018}).

\bibitem{10.1145/3134841}
\bibinfo{author}{Martin, P.}, \bibinfo{author}{Symington, A.} \&
  \bibinfo{author}{Srivastava, M.}
\newblock \bibinfo{journal}{\bibinfo{title}{Slats: Simultaneous localization
  and time synchronization}}.
\newblock {\emph{\JournalTitle{ACM Trans. Cyber-Phys. Syst.}}}
  \textbf{\bibinfo{volume}{2}}, \doiprefix\url{10.1145/3134841}
  (\bibinfo{year}{2018}).

\bibitem{10.1145/3386244}
\bibinfo{author}{Rosales, R.} \& \bibinfo{author}{Paulitsch, M.}
\newblock \bibinfo{journal}{\bibinfo{title}{Composable finite state
  machine-based modeling for quality-of-information-aware cyber-physical
  systems}}.
\newblock {\emph{\JournalTitle{ACM Trans. Cyber-Phys. Syst.}}}
  \textbf{\bibinfo{volume}{5}}, \doiprefix\url{10.1145/3386244}
  (\bibinfo{year}{2021}).

\bibitem{10.1145/3516449}
\bibinfo{author}{Khayatian, M.} \emph{et~al.}
\newblock \bibinfo{journal}{\bibinfo{title}{Plan b: Design methodology for
  cyber-physical systems robust to timing failures}}.
\newblock {\emph{\JournalTitle{ACM Trans. Cyber-Phys. Syst.}}}
  \textbf{\bibinfo{volume}{6}}, \doiprefix\url{10.1145/3516449}
  (\bibinfo{year}{2022}).

\bibitem{10.1145/3078620}
\bibinfo{author}{Krichene, W.}, \bibinfo{author}{Bourguiba, M.~C.},
  \bibinfo{author}{Tlam, K.} \& \bibinfo{author}{Bayen, A.}
\newblock \bibinfo{journal}{\bibinfo{title}{On learning how players learn:
  Estimation of learning dynamics in the routing game}}.
\newblock {\emph{\JournalTitle{ACM Trans. Cyber-Phys. Syst.}}}
  \textbf{\bibinfo{volume}{2}}, \doiprefix\url{10.1145/3078620}
  (\bibinfo{year}{2018}).

\bibitem{10.1145/3384676}
\bibinfo{author}{Bakker, C.}, \bibinfo{author}{Bhattacharya, A.},
  \bibinfo{author}{Chatterjee, S.} \& \bibinfo{author}{Vrabie, D.~L.}
\newblock \bibinfo{journal}{\bibinfo{title}{Hypergames and cyber-physical
  security for control systems}}.
\newblock {\emph{\JournalTitle{ACM Trans. Cyber-Phys. Syst.}}}
  \textbf{\bibinfo{volume}{4}}, \doiprefix\url{10.1145/3384676}
  (\bibinfo{year}{2020}).

\bibitem{10.1145/3439430}
\bibinfo{author}{Bakker, C.}, \bibinfo{author}{Bhattacharya, A.},
  \bibinfo{author}{Chatterjee, S.} \& \bibinfo{author}{Vrabie, D.~L.}
\newblock \bibinfo{journal}{\bibinfo{title}{Metagames and hypergames for
  deception-robust control}}.
\newblock {\emph{\JournalTitle{ACM Trans. Cyber-Phys. Syst.}}}
  \textbf{\bibinfo{volume}{5}}, \doiprefix\url{10.1145/3439430}
  (\bibinfo{year}{2021}).

\bibitem{10.1145/3375716}
\bibinfo{author}{Bellman, K.} \emph{et~al.}
\newblock \bibinfo{journal}{\bibinfo{title}{Self-aware cyber-physical
  systems}}.
\newblock {\emph{\JournalTitle{ACM Trans. Cyber-Phys. Syst.}}}
  \textbf{\bibinfo{volume}{4}}, \doiprefix\url{10.1145/3375716}
  (\bibinfo{year}{2020}).

\bibitem{10.1145/3375403}
\bibinfo{author}{Esterle, L.} \& \bibinfo{author}{Brown, J. N.~A.}
\newblock \bibinfo{journal}{\bibinfo{title}{I think therefore you are: Models
  for interaction in collectives of self-aware cyber-physical systems}}.
\newblock {\emph{\JournalTitle{ACM Trans. Cyber-Phys. Syst.}}}
  \textbf{\bibinfo{volume}{4}}, \doiprefix\url{10.1145/3375403}
  (\bibinfo{year}{2020}).

\bibitem{10.1145/3461669}
\bibinfo{author}{Bakirtzis, G.}, \bibinfo{author}{Fleming, C.~H.} \&
  \bibinfo{author}{Vasilakopoulou, C.}
\newblock \bibinfo{journal}{\bibinfo{title}{Categorical semantics of
  cyber-physical systems theory}}.
\newblock {\emph{\JournalTitle{ACM Trans. Cyber-Phys. Syst.}}}
  \textbf{\bibinfo{volume}{5}}, \doiprefix\url{10.1145/3461669}
  (\bibinfo{year}{2021}).

\end{thebibliography}

\section{Appendix}

\subsection{Adversarial Iterated Boolean Game State Transition Matrices}

\begin{equation}
\begin{aligned}
&    \mathbf{T}_{\text{Unsocial, Optimistic}} = \\
&\begin{bmatrix}
    				0 & \frac{\mathbb{P}(c_0,c_1) + \mathbb{P}(c_1,c_0)}{g_0} & \frac{\mathbb{P}(c_1,c_0)}{g_0} & \frac{\mathbb{P}(c_0\|c_1)}{g_0}& 0 & 0 & 0 & 0 \\
    				0 & 0 & 0 & \frac{\mathbb{P}(c_0,c_1) + \mathbb{P}(c_1,c_0)}{g_1}& 0 & \frac{\mathbb{P}(c_0,c_1)}{g_1}& \frac{\mathbb{P}(c_1,c_0)}{g_1}&  \frac{\mathbb{P}(c_0\|c_1)}{g_1}\\
    				\frac{\mathbb{P}(c_1,c_0)}{g_2}& \frac{\mathbb{P}(c_0\|c_1)}{g_2}&0 &\frac{\mathbb{P}(c_0,c_1)+\mathbb{P}(c_1,c_0)}{g_2} &0 &0 &0 &0  \\
    				0& \frac{\mathbb{P}(c_0,c_1)}{g_3}&\frac{\mathbb{P}(c_1,c_0)}{g_3} &0 &0 &\frac{\mathbb{P}(c_0\|c_1)}{g_3} & 0& \frac{\mathbb{P}(c_0,c_1) + \mathbb{P}(c_1,c_0)}{g_3} \\
    				0& 0& 0& 0& 0& \frac{\mathbb{P}(c_0,c_1)}{g_4}& \frac{\mathbb{P}(c_0,c_1)+\mathbb{P}(c_1,c_0)}{g_4}& \frac{\mathbb{P}(c_0\|c_1)}{g_4} \\
    				0& 0& 0& 0& \frac{\mathbb{P}(c_0,c_1)}{g_5}& 0& \frac{\mathbb{P}(c_0\|c_1)}{g_5}& \frac{2\mathbb{P}(c_0,c_1)+\mathbb{P}(c_1,c_0)}{g_5} \\
    				0& 0& 0& 0& 0& 0& 0& 1 \\
    				0& 0& 0& \frac{\mathbb{P}(c_0\|c_1)+\mathbb{P}(c_1,c_0)}{g_7}& \frac{\mathbb{P}(c_0,c_1)}{g_7} &\frac{\mathbb{P}(c_0,c_1)}{g_7} & \frac{\mathbb{P}(c_1,c_0)}{g_7}&0  \\
                 \end{bmatrix}\\
 &   = \mathbf{T}_{\text{Unsocial, Realistic}}
\end{aligned}
\end{equation}

\begin{equation}
\begin{aligned}
&    \mathbf{T}_{\text{Social, Optimistic}} = \\
& \begin{bmatrix}
    0  & \frac{\mathbb{P}(c_0,c_1)+2\mathbb{P}(c_1,c_0)+\mathbb{P}(c_0\|c_1)}{g_0}  & \frac{2\mathbb{P}(c_0,c_1)}{g_0} & 0  & 0  & 0  & 0  & 0  \\
    \frac{\mathbb{P}(c_0,c_1)}{g_1}   & 0 & \frac{\mathbb{P}(c_0\|c_1)}{g_1}  & \frac{\mathbb{P}(c_1,c_0)}{g_1}  &  0 & 0  & 0  & 0  \\
    0   & \frac{\mathbb{P}(c_0\|c_1)}{g_2}  & 0 & \frac{\mathbb{P}(c_0,c_1)}{g_2}  & 0  & 0  & 0  & 0  \\
    \frac{\mathbb{P}(c_1,c_0)}{g_3}   &  0 & \frac{\mathbb{P}(c_0,c_1)+\mathbb{P}(c_1,c_0)}{g_3}  & 0 & 0  & 0  & 0  & \frac{\mathbb{P}(c_0,c_1) + \mathbb{P}(c_1,c_0) + \mathbb{P}(c_0\|c_1)}{g_3}   \\
    \frac{\mathbb{P}(c_0,c_1)}{g_4}   & 0  & \frac{\mathbb{P}(c_1,c_0)+\mathbb{P}(c_0\|c_1)}{g_4}  &  0 & 0 & 0  & \frac{\mathbb{P}(c_0,c_1)}{g_4}  & 0  \\
     0  &0   & 0  & \frac{\mathbb{P}(c_0,c_1)}{g_5}  & \frac{\mathbb{P}(c_1,c_0)}{g_5}  & 0 & \frac{\mathbb{P}(c_0\|c_1)}{g_5}  & \frac{\mathbb{P}(c_1,c_0)}{g_5}  \\
      0 & 0  & \frac{\mathbb{P}(c_0,c_1)+\mathbb{P}(c_1,c_0)+\mathbb{P}(c_0\|c_1)}{g_6}  & 0  & \frac{\mathbb{P}(c_0,c_1)}{g_6}  & 0  & 0 & 0  \\
      0 & 0  & 0  & 0  & \frac{\mathbb{P}(c_0\|c_1)}{g_7}  & \frac{\mathbb{P}(c_1,c_0)}{g_7}  & \frac{\mathbb{P}(c_0,c_1)+\mathbb{P}(c_1,c_0)}{g_7}  & 0 \\  				
  \end{bmatrix}
\end{aligned}
\end{equation}

\begin{equation}
\begin{aligned}
&    \mathbf{T}_{\text{Social, Realistic}} = \\
& \begin{bmatrix}
    				0&\frac{\mathbb{P}(c_0,c_1)}{g_0}&\frac{\mathbb{P}(c_1,c_0)}{g_0}&\frac{\mathbb{P}(c_0,c_1)+\mathbb{P}(c_0\|c_1)}{g_0}&\frac{\mathbb{P}(c_1,c_0)}{g_0}&0&0&0\\
    				\frac{\mathbb{P}(c_1,c_0)}{g_1}&0&0&\frac{\mathbb{P}(c_0,c_1)+\mathbb{P}(c_1,c_0)}{g_1}&0&\frac{\mathbb{P}(c_0,c_1)}{g_1}&0&\frac{\mathbb{P}(c_0\|c_1)}{g_1}\\
    				\frac{\mathbb{P}(c_0,c_1)+\mathbb{P}(c_1,c_0)}{g_2}&0&0&\frac{\mathbb{P}(c_1,c_0)}{g_2}&\frac{\mathbb{P}(c_0\|c_1)}{g_2}&0&\frac{\mathbb{P}(c_0,c_1)}{g_2}&0\\
    				\frac{\mathbb{P}(c_0\|c_1)}{g_3}&\frac{\mathbb{P}(c_1,c_0)}{g_3}&\frac{\mathbb{P}(c_0,c_1)}{g_3}&0&0&0&0&\frac{\mathbb{P}(c_1,c_0)}{g_3}\\
    				\frac{\mathbb{P}(c_0,c_1)}{g_4}&0&\frac{\mathbb{P}(c_0\|c_1)}{g_4}&0&0&0&\frac{\mathbb{P}(c_0,c_1)+\mathbb{P}(c_1,c_0)}{g_4}&0\\
    				0&0&0&0&\frac{\mathbb{P}(c_0,c_1)}{g_5}&0&\frac{\mathbb{P}(c_0\|c_1)}{g_5}&\frac{\mathbb{P}(c_0,c_1)+\mathbb{P}(c_1,c_0)}{g_5}\\
    				0&0&\frac{\mathbb{P}(c_1,c_0)}{g_6}&0&\frac{\mathbb{P}(c_0,c_1)}{g_6}&0&0&\frac{\mathbb{P}(c_0,c_1)+\mathbb{P}(c_1,c_0)+\mathbb{P}(c_0\|c_1)}{g_6}\\
    				0&0&0&\frac{\mathbb{P}(c_0,c_1)+\mathbb{P}(c_1,c_0)+\mathbb{P}(c_0\|c_1)}{g_7}&0&0&\frac{\mathbb{P}(c_1,c_0)}{g_7}&0\\
                 \end{bmatrix}
\end{aligned}
\end{equation}

\end{document}